\documentclass[review,12pt]{elsarticle}

\usepackage{amssymb}
\usepackage{amsthm}
\usepackage{amsmath}
\usepackage{amssymb}
\usepackage{pifont}

\usepackage{booktabs} 
\usepackage{placeins}
\usepackage{multirow}
\usepackage{mathrsfs}
\usepackage{graphicx}
\usepackage{epstopdf}
\usepackage{float}
\usepackage{caption}
\usepackage{subcaption}
\usepackage{bm}
\usepackage{bbm}
\usepackage{mathrsfs}
\usepackage{cleveref}
\usepackage{soul}
\usepackage{accents}
\usepackage{url} 
\usepackage{graphicx}
\usepackage{xcolor}
\usepackage{courier} 
\usepackage{listings} 
\usepackage{tabu} 
\usepackage{longtable}
\usepackage{changepage} 
\usepackage[margin=2cm]{geometry}
\biboptions{sort&compress} 

\journal{Engineering with Computers}

\makeatletter
\def\@author#1{\g@addto@macro\elsauthors{\normalsize%
    \def\baselinestretch{1}%
    \upshape\authorsep#1\unskip\textsuperscript{%
      \ifx\@fnmark\@empty\else\unskip\sep\@fnmark\let\sep=,\fi
      \ifx\@corref\@empty\else\unskip\sep\@corref\let\sep=,\fi
      }%
    \def\authorsep{\unskip,\space}%
    \global\let\@fnmark\@empty
    \global\let\@corref\@empty  
    \global\let\sep\@empty}%
    \@eadauthor={#1}
}
\makeatother

\begin{document}

\begin{frontmatter}



\title{A computational framework to predict the spreading of Alzheimer's disease}


\author[Uniovi]{Ana Vazquez-Palomo}

\author[Uniovi]{Covadonga Beteg\'{o}n}

\author[Oxf]{Johannes Weickenmeier}

\author[Oxf]{Emilio Mart\'{\i}nez-Pa\~neda\corref{cor1}}
\ead{emilio.martinez-paneda@eng.ox.ac.uk}

\address[Uniovi]{Department of Construction and Manufacturing Engineering, University of Oviedo, Gij\'{o}n 33203, Spain}

\address[Oxf]{Department of Engineering Science, University of Oxford, Oxford OX1 3PJ, UK}

\cortext[cor1]{Corresponding author.}

\begin{abstract}
Alzheimer’s disease is characterised by the spreading of misfolded proteins and progressive structural changes in the brain. Despite significant clinical research, understanding how microscopic protein dynamics translate into macroscopic tissue degeneration remains a major challenge. In this work, we present a three-dimensional, finite element-based computational framework to model disease progression by combining multi-protein transport and brain tissue deformation within anatomically realistic geometries. The propagation of toxic tau and amyloid-$\beta$ proteins is described using reaction–diffusion equations of the Fisher-Kolmogorov type, incorporating prion-like growth mechanisms and anisotropic transport along white matter fibre tracts. Brain atrophy is represented through a hyperelastic constitutive model driven by protein-dependent volume loss. Subject-specific simulations are achieved through an automated preprocessing pipeline that generates finite element meshes and reconstructs axonal orientation fields from medical imaging data. The model reproduces key morphological patterns observed in Alzheimer’s disease and shows good quantitative agreement with longitudinal imaging measurements. Overall, the proposed framework offers an extensible computational platform for studying Alzheimer's disease progression across subject-specific brain geometries. The models developed, including the image processing framework (\texttt{BrainImage2Mesh}) and the coupled bio-chemo-mechanical \texttt{COMSOL} finite element implementation, are made freely available to download at \url{https://mechmat.web.ox.ac.uk/codes}. \\
\medskip
\end{abstract}


\end{frontmatter}

\section*{Nomenclature}

\begin{tabbing}
XXXX \= \kill

$\tilde{a}_0$ \> Healthy A$\beta$ production rate \\
$\tilde{a}_1$ \> Healthy A$\beta$ clearance rate \\
$\tilde{a}_{12}$ \> Healthy to toxic A$\beta$ conversion rate \\
${a}_1$ \> Toxic A$\beta$ clearance rate \\
$\mathbf{a}$ \> Initial axonal unit vector \\
$\mathbf{B}$ \> Body force vector in the reference configuration \\
$\mathbf{b}$ \> Body force vector in the current configuration \\
$\mathcal{B}_0$ \> Initial reference configuration \\
$\mathcal{B}_t$ \> Current configuration \\
$c_i$ \> Protein concentration \\
$c_i^{\text{lim}}$ \> Limit protein concentration \\
$\bar{c}_i$ \> Normalised protein concentration \\
$\bar{c}_{i_0}$ \> Initial normalised toxic protein concentration \\
$\bar{c}_{\text{tau}}^{\text{crit}}$ \> Critical value of the normalised tau concentration \\
$\mathcal{D}$ \> Local dissipation density \\
$\mathcal{D}_{\text{chem}}$ \> Chemical dissipation density \\
$\bm{D}_i$ \> Diffusivity tensor in the reference configuration \\
$\bm{D}_{\mathrm{axn}}$ \> Axonal diffusion tensor in the reference configuration \\
$\bm{D}_{\mathrm{ext}}$ \> Extracellular diffusion tensor in the reference configuration \\
$\bm{d}_i$ \> Diffusivity tensor in the current configuration \\
$\bm{d}_{\mathrm{axn}}$ \> Axonal diffusion tensor in the current configuration \\
$\bm{d}_{\mathrm{ext}}$ \> Extracellular diffusion tensor in the current configuration \\
${\bm{F}}$ \> Deformation gradient tensor \\
$\bm{F}^a$ \> Inelastic deformation gradient tensor \\
$\bm{F}^e$ \> Elastic deformation gradient tensor \\
$G$ \> Second Lamé parameter or shear modulus \\
$G_0$ \> Healthy ageing-related atrophy rate \\
$G_c$ \> Accelerated atrophy rate \\
$\bm{I}$ \> Identity tensor in the reference configuration \\
$\bm{i}$ \> Identity tensor in the current configuration \\
$i$ \> Chemical species index ($i=\text{tau},\,\mathrm{A}\beta$) \\
$J$ \> Deformation Jacobian \\
$J^a$ \> Inelastic deformation Jacobian \\
$J^e$ \> Elastic deformation Jacobian \\
$\tilde{k}_0$ \> Healthy tau production rate \\
$\tilde{k}_1$ \> Healthy tau clearance rate \\
$\tilde{k}_{12}$ \> Healthy to toxic tau conversion rate \\
$\tilde{k}_3$ \> Tau--A$\beta$ coupling rate \\
${k}_1$ \> Toxic tau clearance rate \\
$k_\vartheta$ \> Atrophy stiffness parameter \\
$\bm{M}_i$ \> Mobility tensor \\
$\mathbf{N}$ \> Outward unit normal vector to the boundary in the reference configuration \\
$\mathbf{n}$ \> Outward unit normal vector to the boundary in the current configuration \\
$\bm{P}$ \> First Piola--Kirchhoff stress tensor \\
$\mathcal{P}_{\text{int}}$ \> Internal power density \\
$\mathbf{Q}_i$ \> Transport flux vector in the reference configuration \\
$\mathbf{q}_i$ \> Transport flux vector in the current configuration \\
$R_i$ \> Reaction rate in the reference configuration \\
$r_i$ \> Reaction rate in the current configuration \\
$R_g$ \> Ideal gas constant \\
$t$ \> Time \\
$T$ \> Absolute temperature \\
$\bar{\mathbf T}$ \> Traction vector \\
$\mathbf{u}$ \> Displacement vector \\
$\boldsymbol{\varphi}$ \> Deformation mapping \\
$\mathbf{X}$ \> Position vector in the reference configuration \\
$\mathbf{x}$ \> Position vector in the current configuration \\
$\alpha_{\text{A}\beta}$ \> Global A$\beta$ rate \\
$\alpha_{\text{tau}}$ \> Global tau rate \\
$\gamma$ \> Acceleration of natural atrophy function \\
$\eta_i$ \> Scalar kinematic chemical field \\
$\eta_\vartheta$ \> Viscosity-like atrophy parameter \\
$\kappa$ \> Transition sharpness factor \\
$\lambda$ \> First Lamé parameter \\
$\lambda_a$ \> Stretch along the axonal network \\
$\mu_i$ \> Chemical potential \\
$\psi$ \> Free energy density \\
$\psi^a$ \> Atrophy free energy density \\
$\psi^c$ \> Chemical free energy density \\
$\psi^e$ \> Atrophy-weighted elastic strain free energy density \\
$\psi^e_0$ \> Pristine elastic strain free energy density \\
$\Phi_a$ \> Atrophy dissipation potential \\
$\bm{\sigma}$ \> Cauchy stress tensor \\
$\vartheta$ \> Atrophy factor \\

\end{tabbing}

\section{Introduction}
\label{Introduction}
Alzheimer's disease is a progressive neurodegenerative disorder and the most prevalent form of dementia worldwide. It is characterised by a gradual cognitive and functional decline accompanied by structural changes in the brain \cite{Scheltens2021Lancet}. From a biological perspective, Alzheimer’s disease is commonly described as a progressive process in which misfolded proteins gradually appear and spread throughout the brain. These misfolded proteins, among which amyloid-$\beta$ and tau play a central role, spread by causing healthy proteins to misfold, in a chain reaction akin to that observed in prion diseases. Amyloid-$\beta$ is believed to accumulate first, forming extracellular deposits. Over time, tau proteins become misfolded and begin to spread, mainly following the axonal pathways of the brain \cite{JuckerWalker2013}. The progressive spread of misfolded (toxic) tau proteins is associated with structural neurodegenerative damage, which, on a macroscopic level, is reflected as tissue loss and brain atrophy \cite{Whitwell2008TauAtrophy}. Although the precise causal relationships are still unclear, the simultaneous occurrence of amyloid-$\beta$ and tau indicates that their behaviours are likely to be interconnected. Considering both proteins together is therefore important, as it may help to better understand how their combined dynamics relate to the progression of the disease.\\

Several mathematical models have been developed to describe protein propagation and subsequent structural brain changes, with the ultimate goal of identifying early predictive indicators of neurodegeneration through suitable biomarkers \cite{Budday2020Review}. This is critical, as the initial symptoms of cognitive decline become noticeable only one to two decades after the first pathological abnormalities, and would allow establishing a prognostic timeframe of disease progression, which could be used to inform clinical trials and pharmacological intervention. Early conceptual work by Matthäus \cite{Matthaus2006} compared diffusion-based descriptions with network models to study the spread of prion-like diseases in the brain. Classical reaction–diffusion formulations were later adopted by building on the Fisher–Kolmogorov equations \cite{Fisher1937,KPP1937}, whereby the propagation dynamics of misfolded proteins are captured without explicitly resolving the transport of healthy proteins \cite{Weickenmeier2019JMPS,fornari2019prion,corti2023discontinuous}. This class of models has become popular due to its simplicity and relatively low computational cost, having been extended to incorporate tissue atrophy \cite{weickenmeier2018multiphysics} and used within realistic brain geometries in both 2D \cite{Schafer2019CMA} and (more recently) 3D \cite{Blinkouskaya2021Frontiers,pederzoli2025coupled}. A more detailed class of models, often referred to as heterodimer formulations \cite{antonietti2024discontinuous}, considers two protein configurations: healthy and misfolded. These models allow explicit simulation of phenomena such as protein aggregation, conversion, and clearance, providing a more detailed representation of the underlying protein kinetics, at the cost of considering additional parameters and higher computation times \cite{Corti2024BrainMultiphysics,antonietti2025numerical}. The interplay between amyloid-$\beta$ and tau proteins has recently been considered in the context of heterodimer formulations \cite{Pal2022SciRep}, capturing how amyloid-$\beta$ acts as an accelerating factor in tau propagation. However, the interaction between amyloid-$\beta$ and tau proteins has yet to be modelled in the context of the widely used Fisher–Kolmogorov formulation.\\

Alongside these modelling efforts, the physical properties of white matter and their influence on substance transport have been studied. Specifically, the relationship between tissue porosity and the directional dependence of transport through white matter has been characterised through mathematical derivation and computational simulations in geometries reconstructed from imaging of brain tissue, with experimental validation \cite{Yuan2024}. Building on this, multiscale computational frameworks have been developed to predict how substances spread through the brain, accounting for the local orientation of nerve fibres as obtained from diffusion tensor imaging \cite{Yuan2025}. Although these studies have been developed primarily in the context of fluid and drug transport \cite{Yuan2024review}, the recognition that white matter anisotropy plays a central role in determining the direction and rate of substance transport is equally relevant to the propagation of misfolded proteins, which is known to follow axonal pathways. The anisotropic transport description adopted in our study is informed by this understanding.\\

In this work, we present a bio-chemo-mechanical formulation for Alzheimer's disease that includes the following novel contributions: (i) a coupled description of amyloid-$\beta$ and tau protein spread in the context of a Fisher–Kolmogorov formulation, inspired by heterodimer kinetic models \cite{Pal2022SciRep}; (ii) the combination of multi-protein spreading dynamics with a bio-chemo-mechanical description of atrophy, building upon the single protein species model by Schäfer et al. \cite{Schafer2019CMA}; and (iii) an anisotropic description of protein transport that incorporates a realistic axonal vector field. Moreover, we present a three-dimensional computational framework based on subject-specific brain meshes, in which anatomically relevant regions associated with Alzheimer’s disease are explicitly represented. A key element of this framework is the development of a custom Linux-based preprocessing pipeline (\texttt{BrainImage2Mesh}), which automates medical image processing, brain segmentation, mesh generation, and axonal orientation reconstruction, providing subject-specific inputs for subsequent finite element simulations. The framework is made freely available at \url{https://mechmat.web.ox.ac.uk/codes}, providing the community with, to our knowledge, the first openly shared computational platform for the prediction of protein spreading and atrophy due to Alzheimer's disease.\\ 

The remainder of this paper is structured as follows. In Section \ref{Sec:Theory}, the proposed formulation is presented in a variationally and thermodynamically consistent fashion. Then, in Section \ref{Sec:Framework}, the computational framework is described, spanning the procedures used to generate subject-specific brain meshes, the approach adopted to reconstruct the axonal orientation field, and the finite element implementation of the model within the \texttt{COMSOL} Multiphysics environment. Representative results are given in Section \ref{Sec:Results}, including validation case studies that demonstrate the ability of the model to capture the evolution of the brain in both healthy ageing and Alzheimer’s disease scenarios. Finally, concluding remarks are given in Section \ref{Sec:Concluding remarks}. 

\section{Theory: a multiphysics model of cerebral atrophy}
\label{Sec:Theory}

We proceed to present our theory, which is formulated in a thermodynamically and variationally-consistent fashion and extends the biochemical degradation models for neurodegenerative diseases on which it is based \cite{Weickenmeier2019JMPS,Schafer2019CMA,Pal2022SciRep}. Brain tissue degradation and the associated changes in brain morphology are modelled as an inelastic volume change, using an internal scalar atrophy variable. A suitable, generalised free energy density is defined to capture the coupling between bio-mechanical deformation, bio-chemical transport, and atrophy-induced shrinkage. As detailed below, our formulation captures the following coupled phenomena: (i) the modulation of tau protein transport due to the presence of amyloid-$\beta$ ($\text{A}\beta$) protein; (ii) the acceleration of cerebral atrophy due to tau protein accumulation; (iii) the brain volumetric shrinkage resulting from atrophy and its impact on hyperelastic tissue deformation; and (iv) the role that deformation plays in protein transport.\\

\noindent \textbf{\textit{Notation}}. We use lightface italic letters for scalars, e.g. $\vartheta$, upright bold letters for vectors, e.g. $\mathbf{Q}$, and bold italic letters, such as $\bm{F}$, for second and higher order tensors. In the material frame, the gradient and the divergence are respectively denoted by $\nabla_{\mathbf{X}}$ and $\text{Div}_{\mathbf{X}}$, while their spatial frame counterparts are $\nabla_{\mathbf{x}}$ and $\text{div}_{\mathbf{x}}$. The dyadic product is represented by $\otimes$, and the norm of a vector $v$ is written as $\| \mathbf{v} \|$. 

\subsection{Kinematics and primal fields}
\label{sec:Kinematics}

The primary unknowns of the coupled bio-chemo-mechanical problem are the displacement vector $\mathbf{u}$ and the concentrations of tau and amyloid-$\beta$ proteins, respectively denoted $c_{\text{tau}}$ and $c_{\text{A}\beta}$. In addition, the degree of brain atrophy is described by means of an internal state variable $\vartheta$, which is defined as the remaining healthy volume fraction, varying from 1 (in completely healthy regions) to 0 (in areas where tissue atrophy is complete). A finite strain formulation is adopted, whereby the primal fields are mapped from the initial (reference) configuration to the current configuration at a time $t$. Let $\mathcal{B}_0\subset\mathbb{R}^3$ denote the reference configuration of the brain domain with boundary $\partial\mathcal{B}_0$. Material coordinates are denoted by $\mathbf{X}\in\mathcal{B}_0$ and spatial coordinates by $\mathbf{x}=\boldsymbol{\varphi}(\mathbf{X},t)$ where $\boldsymbol{\varphi}$ is the deformation mapping, which is related to the displacement field by $\mathbf{u}=\boldsymbol{\varphi}-\mathbf{X}$.\\ 

The kinematics of the shrinking brain are described by means of a fully reversible elastic contribution and an inelastic atrophy-induced contribution, with the latter being dependent on the concentration of tau protein ($c_{\text{tau}}$). Thus, assuming a multiplicative decomposition of the total deformation gradient tensor ($\bm{F}$),
\begin{equation} \label{eq:Kinematics}
   \bm{F} = \nabla_{\mathbf{X}} \boldsymbol{\varphi} = \bm{F}^e \cdot \bm{F}^a (c_{\text{tau}})  \,\,\,\,\, \text{with} \,\,\,\,\,\,\,  J=\text{det} \left( \bm{F} \right) = J^e J^a (c_{\text{tau}}),
\end{equation}

\noindent where $\bm{F}^e$ is the elastic part of the deformation gradient tensor and $\bm{F}^a$ its inelastic atrophy counterpart. As shown in Eq. (\ref{eq:Kinematics}), the multiplicative decomposition extends over to the Jacobian $J$, with $J^e=\text{det} \left( \bm{F}^e \right)$ characterising the elastic volume change and $J^a=\text{det} \left( \bm{F}^a \right)$ describing the volume loss associated with tissue atrophy. \\

The kinematics of protein evolution is typically described by the time derivative of the protein concentration $\dot{c}=\text{d}c/\text{d}t$, which in general can take contributions from both diffusion and reactions (protein aggregation), with the latter being typically described by a reaction rate $R$ and the former by the chemical potential $\mu$ or a transport flux vector $\mathbf{Q}$. Following Refs. \cite{cui2022generalised,duda2018phase}, we define a scalar kinematic chemical field $\eta(\mathbf{X},t)$, such that $\dot{\eta} = \mu$ and,
\begin{equation} 
    \eta(\mathbf{X},t)= \int_0^t \mu (\mathbf{X},t) \, \text{d} t.
\end{equation}

\noindent In addition, it will eventually become useful to define a limit concentration (protein solubility) $c^{\text{lim}}$, below which the protein remains dissolved in a stable, homogeneous solution. This enables defining a normalised concentration $\bar{c}=c/c^{\text{lim}}$. It is important to emphasise that, throughout this manuscript, the concentration variable $c (\mathbf{X},t)$ denotes molar concentration per unit \emph{reference} volume.

\subsection{Virtual power and balance laws}
\label{sec:PVW}

For arbitrary virtual fields $\delta \boldsymbol{\varphi}$, $\delta \eta_{\text{tau}}$ and $\delta \eta_{\text{A}\beta}$, the virtual power equality reads:
\begin{align}
\label{eq:Intvirtual_power}
\int_{\mathcal{B}_0}
\Big( \bm{P} :  \nabla_{\mathbf{X}} \left(\delta \dot{\boldsymbol{\varphi}}  \right) & 
+\sum_{i\in\{\text{tau},\text{A}\beta\}}\big[(\dot c_i - R_i)\,\delta\dot{\eta}_i - \mathbf Q_i\!\cdot\!\nabla_{\mathbf{X}} (\delta\dot{\eta}_i)\big]\Big) \, \text{d}V =  \nonumber \\
& \int_{\mathcal B_0} \mathbf B \cdot \delta \dot{\boldsymbol{\varphi}} \, \text{d}V + \int_{\partial\mathcal B_{0}} \bar{\mathbf T} \cdot \delta \dot{\boldsymbol{\varphi}} \, \text{d}S  - \sum_{i}\int_{\partial\mathcal B_0}\bar q_i\,\delta \dot{\eta}_i\,\mathrm dS,
\end{align}

\noindent where $\bm{P}$ is the first Piola-Kirchhoff stress tensor (work-conjugate to $\bm{F}$), $\mathbf{B}$ is a body force vector, $\bar{\mathbf T}$ is a prescribed traction acting on $\partial\mathcal B_{0}$, and $q$ is a prescribed (outward) biochemical flux acting normal to $\partial\mathcal B_{0}$. In this work, the flux $\mathbf{Q}$ is defined as positive when directed outward. \\

Consider first the mechanical terms in (\ref{eq:Intvirtual_power}). As shown in the Supplementary Material, using the divergence theorem and considering the fundamental lemma of the calculus of variations, one reaches the following balance equation in $\mathcal{B}_0$
\begin{equation} \label{eq:MechEqui}
    \text{Div}_{\mathbf{X}} \,  \bm{P}  + \mathbf{B} = 0, 
\end{equation}

\noindent with natural boundary condition $\bm{P} \cdot \mathbf{N} = \bar{\mathbf T}$, where $\mathbf{N}$ is the outward unit normal vector to the boundary in the reference configuration. One can readily derive the mechanical equilibrium equation in the current configuration $\mathcal{B}_t$ by considering the relationship between the Piola-Kirchhoff and Cauchy stresses, $\bm{P} = J \bm{\sigma} \bm{F}^{-T}$, and Nanson's formula, $\mathbf{n} \, \text{d} S= J \bm{F}^{-T}  \mathbf{N} \, \text{d} S_0$; such that
\begin{equation} \label{eq:MechEqui_t}
    \text{div}_{\mathbf{x}} \,  \bm{\sigma} + \mathbf{b} = 0, 
\end{equation}

\noindent with natural boundary condition $\bm{\sigma} \cdot \mathbf{n} = \mathbf{t}$ on $\partial \mathcal{B}_t$.\\ 

By following a similar protocol (see the Supplementary Material), the balance laws for the biochemical problem can be derived as,
\begin{equation}\label{eq:EQtau1}
    \dot{c}_{\text{tau}} +   \text{Div}_{\mathbf{X}} \,  \mathbf{Q}_{\text{tau}}  = R_{\text{tau}},
\end{equation}
\begin{equation}\label{eq:EQabeta1}
   \dot{c}_{\mathrm{A}\beta} +  \text{Div}_{\mathbf{X}} \,  \mathbf{Q}_{\mathrm{A}\beta}  = R_{\mathrm{A}\beta} 
\end{equation}

\noindent in $\mathcal{B}_0$, with natural boundary conditions $\bar q_{\text{tau}} = \mathbf{Q}_{\text{tau}} \cdot \mathbf{N} $ and $\bar q_{\mathrm{A}\beta} = \mathbf{Q}_{\mathrm{A}\beta} \cdot \mathbf{N} $, on $\partial\mathcal B_{0}$. For convenience, we choose to work always with the material concentration as the primary variable for the transport problem. Accordingly, the balance law in the current configuration can be readily derived (as shown in the Supplementary Material) such that, for any choice of protein $i$, one has
\begin{equation}\label{eq:EQtransportBt}
   \dot{c}_i +  \text{div}_{\mathbf{x}} \, \left( J \mathbf{q}_i \right)   = R_i 
\end{equation}

\noindent in $\mathcal{B}_t$, which can also be formulated in terms of the spatial source term, upon noting that $r_i = R_i/J$. 

\subsection{Energy imbalance}
\label{sec:EnergyImbalance}

A generalised bio-chemo-mechanical free energy density $\psi \left(\mathbf{X}, t \right)$ is formulated, ensuring consistency with continuum thermodynamics \cite{gurtin2010mechanics}. To that end, let us define the internal power density $\mathcal{P}_{\text{int}}$, considering the virtual power expression (\ref{eq:Intvirtual_power}), replacing the virtual fields ($\delta \dot{\boldsymbol{\varphi}}$, $\delta \dot{\eta}$) by realisable velocity fields ($\dot{\boldsymbol{\varphi}}$, $\mu$), and separating the contribution from each protein species, such that:
\begin{equation}
    \mathcal{P}_{\text{int}} = \bm{P} : \nabla_{\mathbf{X}} \dot{\varphi} + \left( \dot{c}_{\text{tau}} - R_{\text{tau}} \right) \mu_{\text{tau}} + \left( \dot{c}_{\text{A} \beta} - R_{\text{A} \beta} \right) \mu_{\text{A} \beta} - \mathbf{Q}_{\text{tau}} \cdot \nabla_{\mathbf{X}} \mu_{\text{tau}} - \mathbf{Q}_{\text{A} \beta} \cdot \nabla_{\mathbf{X}} \mu_{\text{A} \beta}.
\end{equation}

This internal power density $\mathcal{P}_{\text{int}}$ is related to the local dissipation density $\mathcal{D}$ and the rate of change of the stored free energy, $\dot{\Psi}$ as $\mathcal{D}=\mathcal{P}_{\text{int}}-\dot{\psi}$. Since the second law of thermodynamics requires that the dissipation be non-negative for all admissible processes ($\mathcal{D} \geq 0$), a thermodynamically-consistent definition of $\psi$ must satisfy the following local energy imbalance,
\begin{equation}\label{eq:2ndlaw}
    \dot{\psi} \leq \bm{P} : \nabla_{\mathbf{X}} \dot{\varphi} + \mu_{\text{tau}} \dot{c}_{\text{tau}} + \mu_{\text{A} \beta} \dot{c}_{\text{A} \beta} - \mathbf{Q}_{\text{tau}} \cdot \nabla_{\mathbf{X}} \mu_{\text{tau}} - \mathbf{Q}_{\text{A} \beta} \cdot \nabla_{\mathbf{X}} \mu_{\text{A} \beta} - R_{\text{tau}} \mu_{\text{tau}} -  R_{\text{A} \beta}  \mu_{\text{A} \beta}.
\end{equation}

\subsection{Free energy density}
\label{sec:FreeEnergy}

We proceed to define the functional form of the free energy density in (\ref{eq:2ndlaw}). First, we decompose $\psi$ into three terms: (i) one associated with the elastic deformation of the brain ($\psi^e$), considering the impact of atrophy; (ii) a chemical term ($\psi^c$), which accounts for the evolution of toxic tau and amyloid-$\beta$ proteins; and, (iii) an atrophy term that captures the interactions between brain shrinkage and bio-chemistry. Hence,
\begin{equation}
    \psi \left( \bm{F}^e, c_{\text{tau}}, c_{\text{A} \beta}, \vartheta \right) = \psi^e \left( \bm{F}^e, \vartheta \right) + \psi^c \left(  c_{\text{tau}} , c_{\text{A} \beta} \right) +  
    \psi^a \left(  \vartheta, c_{\text{tau}} \right).
\end{equation}

The elastic term follows a compressible neo-Hookean hyperelastic model, which is well-suited for soft biological tissues undergoing large deformations. In addition, it is scaled by the atrophy factor $\vartheta$, which we define to be equal to $J^a$ below. Accordingly,
\begin{equation}
\psi^e =\vartheta \psi^e_0 = \vartheta \left\{ \frac{G}{2} \left[ \bm{F}^e : \bm{F}^e - 3 - 2 \ln(J^e) \right] + \frac{\lambda}{2}  \ln^2(J^e) \right\},
\end{equation}
where $J^e=\text{det} \bm{F}^e$, $\psi^e_0$ is the pristine elastic strain energy density, and $G$ and $\lambda$ are the Lamé parameters.\\

The chemical free energy density is defined as follows,
\begin{equation}
    \psi^c = 
   \sum_{i\in\{\text{tau},\text{A}\beta\}}  \mu^0_{i} c_i + R_g T \left[ c_i \ln \left( \frac{c_i}{c^{\text{lim}}_i} \right) - c_i  \right],
\end{equation}

\noindent with $R_g$ being the ideal gas constant and $T$ the absolute temperature. As shown below, this choice renders the ideal dilute chemical potential and defers to the constitutive definitions the influence of atrophy, anisotropy, and tau-$\text{A}\beta$ protein interactions. Alternatively, one could define the chemical free energy as a function of the atrophy factor $\vartheta$ and/or a coupling term involving $c_{\text{tau}}$ and $c_{\text{A} \beta}$; while variationally and thermodynamically consistent, this introduces cross-couplings that are not present in existing theories (e.g., an influence of atrophy on the flux, via $\nabla \mu$, or an influence of $c_{\text{A} \beta}$ on the transport of $c_{\text{tau}}$). \\

Finally, the atrophy term is given as,
\begin{equation}
    \psi^a = \frac{k_\vartheta}{2} \left( \vartheta - 1 \right)^2, 
\end{equation}

\noindent with the atrophy factor $\vartheta$ evolving based on $c_{\text{tau}}$ and $k_\vartheta>0$ being a penalty-like atrophy stiffness parameter, which allows us to regularise the atrophy response, as discussed in the constitutive section below.

\subsection{Constitutive theory}
\label{sec:Constitutive}

We proceed to make appropriate constitutive choices for the bio-chemo-mechanical processes involved, ensuring thermodynamic consistency. 

\subsubsection{Finite hyperelasticity of a shrieking brain}

The derivation of the Piola-Kirchhoff stress tensor follows the elastic strain energy density definition as,
\begin{equation}
    \bm{P}= \frac{\partial \psi^e}{\partial \bm{F}} = \left( \frac{\partial \psi^e}{\partial \bm{F}^e}  \right) \left( \bm{F}^{a} \right)^{-T} = \vartheta \left[ G \left( \bm{F}^e - \left( \bm{F}^{e} \right)^{-T} \right) + \lambda \ln J^e \left( \bm{F}^{e} \right)^{-T} \right] \left( \bm{F}^{a} \right)^{-T},
\end{equation}

\noindent resulting in its scaling by the atrophy factor $\vartheta$, which provides a direct coupling between volumetric shrinkage and the load-carrying capacity of the brain tissue. Accordingly, the Cauchy stress tensor reads,
\begin{equation}
    \bm{\sigma} = \vartheta \left[ \frac{G}{J} \left( \bm{F}^e \left( \bm{F}^{e} \right)^{T}   - \bm{I} \right) + \frac{\lambda \ln J^e}{J} \bm{I} \right]
\end{equation}

\subsubsection{Protein spread}

For any choice of protein $i$, the chemical potential can be readily derived from the chemical free energy density as,
\begin{equation}
    \mu_{i} = \frac{\partial \psi^c}{\partial c_{i}} =   \mu^0_{i} + RT \ln \left( \frac{c_i}{c^{\text{lim}}_i} \right)
\end{equation}

Then, as standard, the flux is assumed to be related to the chemical potential through a linear, Onsager-type relationship,
\begin{equation}\label{eq:Flux1}
    \mathbf{Q}_i = - \bm{M}_i \cdot \nabla_{\mathbf{X}} \, \mu_i
\end{equation}

\noindent where $\bm{M}$ is a mobility tensor. To recover the characteristic reaction-diffusion equations of protein transport, the following non-linear mobility tensor is defined,
\begin{equation}\label{eq:Mobility}
    \bm{M}_i (c_i, \bm{F}  ) =  \frac{c_i \bm{D}_i (\bm{F} )}{RT}  
\end{equation}

\noindent where $\bm{D}_i$ is a protein-specific, diffusivity tensor. As shown in the Supplementary Material, substituting (\ref{eq:Mobility}) into (\ref{eq:Flux1}), one obtains the following Fickian-like flux,
\begin{equation}\label{eq:FickFlux}
    \mathbf{Q}_i = - \bm{D}_i \nabla_{\mathbf{X}} \, c_i.
\end{equation}

These choices ensure thermodynamic consistency, as $\bm{D}$ is positive definite and $c>0$, which implies that the chemical term of the dissipation density fulfils:
\begin{equation}
    \mathcal{D}_{\text{chem}} = - \mathbf{Q}_i \cdot \nabla_{\mathbf{X}} \mu = \frac{R_gT}{c_i} \left( \nabla c_i \right)^T \bm{D} \left( \nabla c_i \right) \geq 0.
\end{equation}

\noindent We shall now particularise these choices for the specific cases of misfolded tau and amyloid-$\beta$ ($\text{A}\beta$) proteins.\\

\noindent \textbf{Amyloid-$\beta$ ($\text{A}\beta$) protein}. The material diffusion tensor describing the transport of $\text{A}\beta$ proteins is given as,
\begin{equation}
    \bm{D}_{\text{A}\beta} = D^{\text{ext}} \bm{I}
    \label{eq:dif_ab1},
\end{equation}

\noindent where $D_{ext}$ accounts for extracellular diffusion, which takes place in both grey and white matter. Since amyloid-$\beta$ propagates exclusively through the extracellular space, axonal transport ($d_{axn}$) is not considered in this case, resulting in an isotropic diffusion process (unlike the tau protein case, discussed below). The spatial counterpart to $\bm{D}_{\text{A}\beta}$, denoted $\bm{d}_{\text{A}\beta}$, can be readily obtained by pushing forward the material diffusion tensor to the spatial configuration: $\bm{d}_{\text{A}\beta}=J^{-1} \bm{F}\, \bm{D}_{\text{A}\beta}\, \bm{F}^{\mathrm T}=J^{-1} D^{\text{ext}} \bm{F}\bm{F^\text{T}}$. However, we chose to neglect the influence of volumetric deformation on extracellular diffusion. Accordingly, the diffusion tensor is directly defined in the spatial configuration as:
\begin{equation}
    \bm{d}_{\text{A}\beta} = d^{\text{ext}} \bm{i},
    \label{eq:dif_ab}
\end{equation}

\noindent where $\bm i$ denotes the spatial identity tensor and $d_{\mathrm{ext}}$ the extracellular diffusion coefficient defined in the current configuration.\\

\begin{figure}[h]
    \centering
    \includegraphics[width=0.9\textwidth]{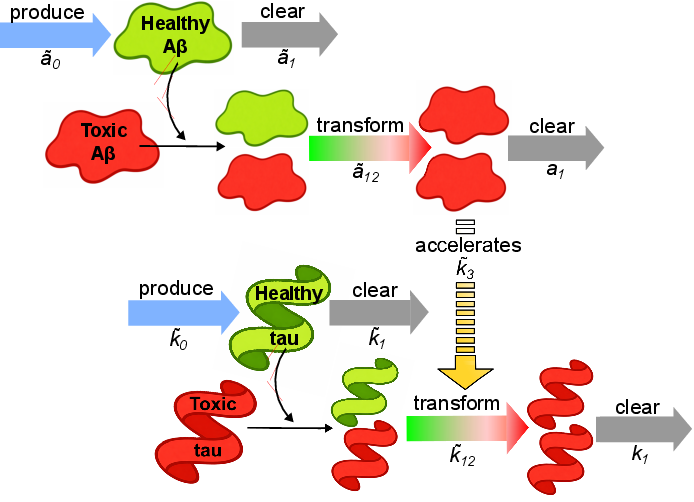}
    \caption{Kinetics of healthy and misfolded tau and amyloid-$\beta$ (A$\beta$) proteins. The diagram illustrates the production (blue arrows), clearance (grey arrows) and prion-like conversion (green/red bi-colour arrows) of healthy proteins into toxic species, with each step being described by a relevant reaction rate parameter. The role of misfolded A$\beta$ protein in accelerating the spreading of misfolded tau proteins is highlighted with a yellow discontinuous arrow, which links the two protein transport schemes.}
    \label{fig:protein_spread}
\end{figure}

With the flux contribution defined, let us now turn our attention to the source term of the protein balance, Eq. (\ref{eq:EQabeta1}). Following the work by Schäfer et al.~\cite{Schafer2019CMA}, the source term is defined by starting from a heterodimer kinetic model and adopting a series of simplifications that allow us to capture the underlying kinetics purely as a function of the concentration of misfolded (toxic) amyloid-$\beta$ protein $c_{\text{A} \beta}$. Thus, consider the conversion kinetics described in Fig. \ref{fig:protein_spread}. The concentration of healthy amyloid-$\beta$ protein $\tilde{c}_{\text{A} \beta}$ is described by a production term (with rate constant $\tilde{a}_0$), a clearance term (with rate constant $\tilde{a}_1$), and a conversion to toxic protein term (with rate constant $\tilde{a}_{12}$), such that,
\begin{equation}
\frac{\partial \tilde{c}_{\text{A}\beta}}{\partial t}
=
\tilde{a}_0
- \tilde{a}_1 \tilde{c}_{\text{A}\beta}
- \tilde{a}_{12} c_{\text{A}\beta} \tilde{c}_{\text{A}\beta},
\label{eq:hetero_abeta_healthy}.
\end{equation}

Initially, the amount of healthy protein is much larger than that of misfolded protein ($\tilde{c}_{\text{A}\beta} >> c_{\text{A}\beta}$) and $\partial \tilde{c}_{\text{A}\beta} / \partial t \approx 0$ \cite{Schafer2019CMA}, such that $\tilde{c}_{\text{A}\beta} \approx \tilde{a}_0 / (\tilde{a}_1+\tilde{a}_{12} c_{\text{A}\beta})$. This can be further approximated using a Taylor series evaluated at $\tilde{a}_{12} c_{\text{A}\beta}/ \tilde{a}_1=0$, rendering,
\begin{equation}\label{eq:c_Absimple}
    \tilde{c}_{\text{A}\beta} \approx \frac{\tilde{a}_0}{\tilde{a}_1} \left( 1 -  \frac{\tilde{a}_{12}}{\tilde{a}_1}c_{\text{A}\beta}\right).
\end{equation}

Now, let us define the evolution of the misfolded concentration $c_{\text{A}\beta}$. As per Fig. \ref{fig:protein_spread}, it must involve a clearance term (with rate $a_1$) and a conversion term, of opposite sign to that in (\ref{eq:hetero_abeta_healthy}); accordingly,
\begin{equation}
\frac{\partial c_{\text{A}\beta}}{\partial t}
=
- a_1 c_{\text{A}\beta}
+ \tilde{a}_{12} c_{\text{A}\beta} \tilde{c}_{\text{A}\beta}.
\label{eq:hetero_abeta_misfolded} 
\end{equation}

Substituting (\ref{eq:c_Absimple}) into (\ref{eq:hetero_abeta_misfolded}) one reaches,
\begin{equation}
 \frac{\partial c_{\text{A}\beta}}{\partial t}
=  \left( \tilde{a}_{12}  \frac{\tilde{a}_0}{\tilde{a}_1} - a_1 \right) c_{\text{A}\beta} - \frac{\tilde{a}_{12}^2 \tilde{a}_{0}}{\tilde{a}_{1}^2} c_{\text{A}\beta}^2,
\end{equation}

\noindent and by re-arranging,
\begin{equation}\label{eq:cABeta_hetero}
 \frac{\partial c_{\text{A}\beta}}{\partial t}
=  \left( \tilde{a}_{12}  \frac{\tilde{a}_0}{\tilde{a}_1} - a_1 \right)  c_{\text{A}\beta} \left( 1 - \frac{\frac{\tilde{a}_{12}^2 \tilde{a}_{0}}{\tilde{a}_{1}^2}}{\tilde{a}_{12}  \frac{\tilde{a}_0}{\tilde{a}_1} - a_1} c_{\text{A}\beta} \right) = \alpha_{\text{A}\beta} c_{\text{A}\beta} \left(1 - \frac{c_{\text{A}\beta}}{c_{\text{A}\beta}^{\text{lim}}} \right),
\end{equation}

\noindent we can identify a global growth-rate parameter $\alpha_{\text{A}\beta}=\tilde{a}_{12} \tilde{a}_{0}/\tilde{a}_1 - a_1$, associated with amyloid-$\beta$ dynamics, and the limit (maximum) concentration of amyloid-$\beta$ protein,
\begin{equation}
   c_{\text{A}\beta}^{\text{lim}}=\frac{\tilde{a}_1}{\tilde{a}_{12}} \left( 1 - \frac{\tilde{a}_1 a_1}{\tilde{a}_{12} \tilde{a}_0} \right) 
\end{equation}

Accordingly, to recover (\ref{eq:cABeta_hetero}), the source term in (\ref{eq:EQtransportBt}) describing the local growth of toxic amyloid-$\beta$ protein must be given by,
\begin{equation}\label{eq:Rab}
    R_{\text{A}\beta} \left( c_{\text{A} \beta} \right) =  \alpha_{\text{A}\beta} c_{\text{A}\beta} \left(1 - \frac{c_{\text{A}\beta}}{c_{\text{A}\beta}^{\text{lim}}} \right).
\end{equation}

Then, considering Eqs. (\ref{eq:EQabeta1}), (\ref{eq:EQtransportBt}), (\ref{eq:FickFlux}), and (\ref{eq:Rab}) one can formulate the misfolded A$\beta$ protein transport balance equation in $\mathcal B_t$, accounting for the appropriate constitutive choices:
\begin{equation}\label{eq:AbProt}
\frac{\partial c_{\text{A}\beta}}{\partial t}= \text{div} \,(\bm{d}_{\text{A}\beta} \nabla c_{\text{A}\beta}) + \alpha_{\text{A}\beta} c_{\text{A}\beta} \left(1 - \frac{c_{\text{A}\beta}}{c_{\text{A}\beta}^{\text{lim}}} \right).
\end{equation}

\noindent \textbf{Tau protein}. The transport of tau proteins is described by a diffusion tensor, which in the material (reference) configuration reads
\begin{equation}
    \bm{D}_\text{tau} = D^{\text{ext}} \bm{I} + D^{\text{axn}} \mathbf{a} \otimes \mathbf{a},
\label{eq:dif_tau_material}
\end{equation}

\noindent where $D_{ext}$ accounts for extracellular diffusion, which takes place in both grey and white matter, and $D_{axn}$ represents transport along axonal fibre, which is only present in white matter. The vector $\mathbf{a}$ denotes the initial axonal unit vector in the material configuration. This is a local quantity that is defined as an input to the model at the nodal level, directly from the interpretation of medical (DTI) images, as elaborated below. To describe tau protein transport in the spatial configuration, the material diffusion tensor can be pushed forward using the deformation gradient, $\bm{d}_\text{tau}= J \bm{F}\, \bm{D}_\text{tau}\, \bm{F}^{\mathrm T}$. Such that, considering Eq. (\ref{eq:dif_tau_material}), the spatial diffusion tensor reads,
\begin{equation}
\bm{d}_\text{tau}=D^{\text{ext}} \bm{F}\bm{F^\text{T}}+D^{\text{axn}}{\bm{F}\cdot\mathbf{a} \otimes \bm{F}\cdot\mathbf{a}}.
\label{eq:dif_tau_spatial}
\end{equation}

Eq. (\ref{eq:dif_tau_spatial}) showcases the interplay between biomechanical deformation, as accounted for \textit{via} $\bm{F}$, and protein transport. Here, for simplicity, extracellular diffusion is assumed to be independent of deformation and thus described by a spatial diffusion coefficient $d^{\text{ext}}$. Also, we chose to present the axonal term normalised by the deformation stretch along the axonal direction $\lambda_a = \|\bm F\cdot \mathbf{a}\|$. Accordingly, the spatial diffusion tensor adopted here reads,
\begin{equation}\label{eq:d_tauFinal}
    \bm{d}_\text{tau}=d^{\text{ext}}\bm{i}+d^{\text{axn}} \frac{\bm{F} \cdot \mathbf{a} \otimes \bm{F} \cdot \mathbf{a}}{\lambda_a^2},
\end{equation}

\noindent where $d_{\mathrm{ext}}$ and $d_{\mathrm{axn}}$ are the extracellular and axonal spatial diffusion coefficients, respectively. These coefficients characterise the intensity of protein transport and are defined in the current configuration, while being assumed invariant under tissue deformation.\\

It remains to define the source term of the balance equation for $c_{\text{tau}}$, which must account not only for the kinetics of healthy and misfolded tau proteins but also for their interaction with the content of misfolded amyloid-$\beta$ ($\text{A}\beta$) protein, as described in Fig. \ref{fig:protein_spread}. Following a similar approach to that adopted in the amyloid-$\beta$ protein case, we start from a heterodimer system of rate equations describing the evolution of healthy $\tilde{c}_{\text{tau}}$ and misfolded (toxic) $c_{\text{tau}}$ tau protein species and then proceed to simplify these to end up with a Fisher-Kolmogorov-like model. Thus, following Ref. \cite{Pal2022SciRep}, the production, clearance and conversion kinetics of healthy and toxic tau proteins are respectively described by,
\begin{equation}
\frac{\partial \tilde{c}_{\text{tau}}}{\partial t}
=
\tilde{k}_0
- \tilde{k}_1 \tilde{c}_{{\text{tau}}}
- \tilde{k}_3 c_{{\text{tau}}} \tilde{c}_{{\text{tau}}} c_{\text{A}\beta}
- \tilde{k}_{12} c_{{\text{tau}}} \tilde{c}_{{\text{tau}}},
\label{eq:hetero_tau_healthy}
\end{equation}
\begin{equation}
\frac{\partial c_{\text{tau}}}{\partial t}
=
- k_1 c_{\text{tau}}
+ \tilde{k}_3 c_{\text{tau}} \tilde{c}_{\text{tau}} c_{\text{A}\beta}
+ \tilde{k}_{12} c_{\text{tau}} \tilde{c}_{\text{tau}},
\label{eq:hetero_tau_misfolded}
\end{equation}

\noindent with $\tilde{k}_0$ denoting the production rate of healthy tau protein, $\tilde{k}_1$ and $k_1$ are the clearance rates of healthy and misfolded tau proteins, respectively, $\tilde{k}_{12}$ represents the conversion rate from healthy to misfolded (toxic) tau, and $\tilde{k}_3$ characterises the kinetics of the coupling term between both proteins. Eqs. (\ref{eq:hetero_tau_healthy})-(\ref{eq:hetero_tau_misfolded}) introduce a unidirectional coupling in which amyloid-$\beta$ enhances tau pathology. A possible extension of the present formulation would be to account for a bidirectional interaction between amyloid-$\beta$ and tau, as well as the inclusion of multiple aggregation states, as in \cite{fornari2020spatially,FranchiHeidaLorenzani2019Arxiv,FranchiLorenzani2017}. Such an extension would be consistent with experimental observations \cite{Bennett2017EnhancedTau,Busche2020Synergy,Roda2022Crosstalk}.\\

As done for the A$\beta$ case, and detailed in the Supplementary Material, the tau rate equations are reduced under the assumption of approximately constant healthy protein concentration, leading to a Fisher-Kolmogorov-type formulation. As a result, the spatial reaction term governing tau aggregation is defined as:
\begin{equation}\label{eq:Rtau}
    R_{\text{tau}} \left( c_{\text{tau}}, c_{\text{A} \beta} \right)  =\alpha_{\text{tau}} \left( \bar{c}_{\text{A} \beta} \right) c_{\text{tau}} \left(1-\frac{c_{\text{tau}}}{c_{\text{tau}}^{\text{lim}}} \right) ,
\end{equation}

\noindent with the following global growth rate parameter, 
\begin{equation}
    \alpha_{\text{tau}} \left( \bar{c}_{\text{A} \beta} \right) =  \frac{\tilde{k}_0}{\tilde{k}_1}\left[\tilde{k}_3 \frac{c_{\text{A}\beta}}{c_{\text{A}\beta}^{\text{lim}}}\frac{\tilde{a}_1}{\tilde{a}_{12}}\left(1 - \frac{\tilde{a}_1 a_1}{\tilde{a}_0 \tilde{a}_{12}}\right)+ \tilde{k}_{12}\right]- k_1 ,
    \label{eq:alpha_tau}
\end{equation}

\noindent which combines tau and A$\beta$ spread dynamics, and a maximum (limit) tau concentration,
\begin{equation}
    c_{\text{tau}}^{\text{lim}} =\frac{\tilde{k}_1}{\tilde{k}_3 {c}_{\text{A}\beta} + \tilde{k}_{12}}
\left(
1
-
\frac{{k}_1 \tilde{k}_1}{\tilde{k}_0 (\tilde{k}_3 {c}_{\text{A}\beta} + \tilde{k}_{12})}
\right).
\end{equation}

Accordingly, the spatial transport equation for tau reads:
\begin{equation} \label{eq:ctau}
    \frac{\partial c_\text{tau}}{\partial t}=\text{div} \, (\,\bm{d}_\text{tau} \nabla_{\mathbf{x}} c_\text{tau}) + \alpha_\text{tau} (\bar{c}_{\text{A} \beta}) c_\text{tau} \left(1-\frac{c_{\text{tau}}}{c_{\text{tau}}^{\text{lim}}} \right).
\end{equation}

\subsubsection{Cerebral atrophy}

Following Sch{\"a}fer \textit{et al.} \cite{Schafer2019CMA}, atrophy is incorporated through the inelastic deformation gradient $\bm{F}^a$. In the simplest case of grey matter, which is typically assumed to be purely isotropic, the atrophy tensor reads,
\begin{equation}
\bm{F}^{a} = \vartheta^{1/3} \, \bm{I}.
\end{equation}

\noindent Here, the atrophy factor $\vartheta \in [1,0]$ describes the remaining healthy volume fraction. Hence, this is purely a geometry quantity and, in both grey and white matter, it can be readily shown that $\vartheta=J^a$ since $J^a = \text{det} \, \bm{F}^a = \vartheta$. Although several biological mechanisms contribute to brain atrophy, the main clinically observable consequence is volume loss. Since the aim of the model is to reproduce structural changes measurable in MRI data rather than to explicitly resolve each microscopic process, the evolution of atrophy is therefore represented by the single scalar variable $\vartheta$, which captures the spatial and temporal evolution of brain volume loss. This simplified representation enables capturing overall volumetric changes but does not explicitly resolve regional differences associated with specific neurodegenerative mechanisms.\\ 

In the case of white matter, atrophy is anisotropic due to the alignment of axons (nerve fibres) along coherent directions (tracts) \cite{Budday2015}. Hence, white matter degeneration is associated with thinning of cortical fiber tracts orthogonal to the fiber direction rather than shrinking evenly in all directions. Accordingly, defining the local axonal direction as $\mathbf{a}$, the atrophy tensor in white matter can be defined as, 
\begin{equation}
\bm{F}^{a} = \sqrt{\vartheta} \, \bm{I} + (1 - \sqrt{\vartheta}) \, \mathbf{a} \otimes \mathbf{a},
\end{equation}

\noindent with $\mathbf{a}$ corresponding to the principal eigenvector, as determined from diffusion tensor imaging (DTI), which captures the directionality and magnitude of diffusion through the eigenvalues and eigenvectors of the diffusion tensor - the eigenvector of the highest eigenvalue is assigned to $\mathbf{a}$ in the computational model, as a pre-processing step. \\

\begin{figure}[h]
    \centering
    \includegraphics[width=0.7\linewidth]{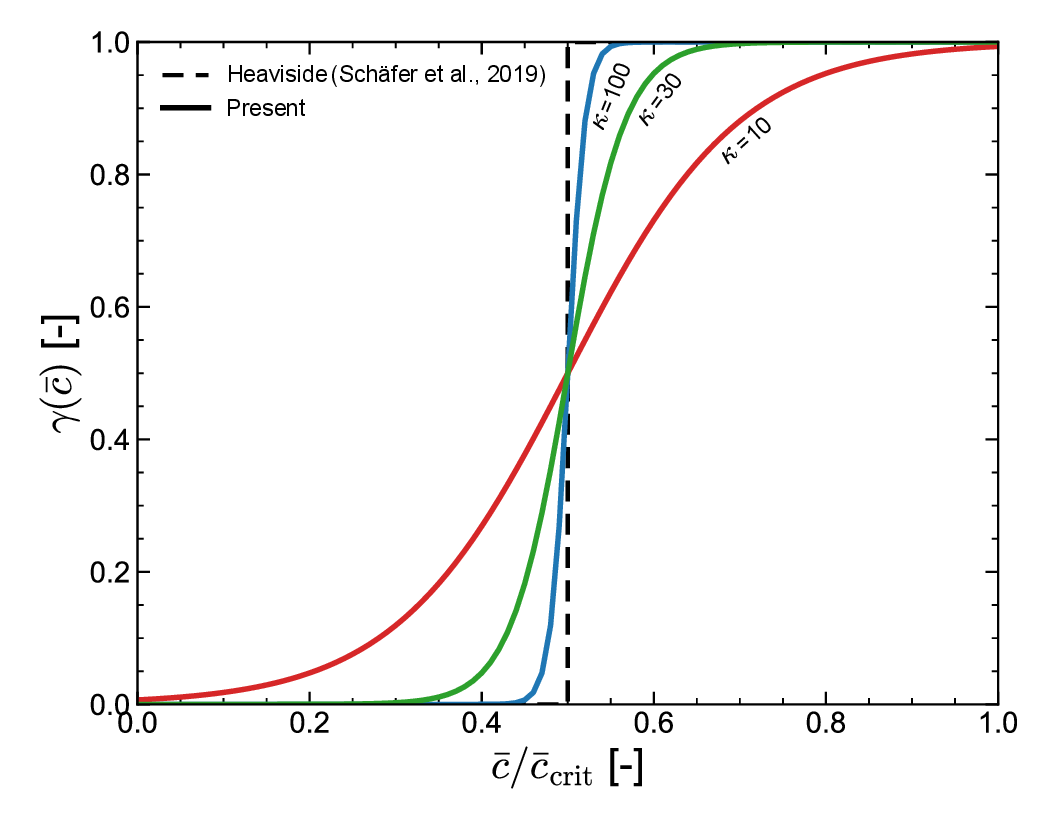}
    \caption{A smooth variation of the atrophy rate degradation function $\gamma(c)$ is proposed, which approximates the Heaviside function-based definition by Sch{\"a}fer \textit{et al.} \cite{Schafer2019CMA}. The parameter $\kappa$ governs the transition; a value of $\kappa=100$ is chosen here to facilitate numerical analysis while capturing the notable jump in atrophy rate expected when the toxic tau concentration reaches a critical quantity.}
    \label{fig:gammaccrit}
\end{figure}

To define thermodynamically-consistent constitutive choices for the atrophy evolution, and its interplay with misfolded tau protein concentration, let us first define an atrophy dissipation potential as,
\begin{equation}
    \Phi_a ( \dot{\vartheta} ) = \frac{\eta_\vartheta}{2} \left[ \dot{\vartheta} -  \left( 1 + \gamma (c_{\text{tau}} ) G_0 \right) \right]^2,
\end{equation}

\noindent with $G_0$ being a negative quantity that characterises the healthy ageing-related rate and $\gamma (c_{\text{tau}} )$ a function that describes the acceleration of natural atrophy due to an increase in misfolded tau protein concentration. We also define $\eta_\vartheta > 0$ as a viscosity-like atrophy parameter. Accordingly, the force balance ($\partial_{\dot{\vartheta}} \Phi + \partial_\vartheta \psi^a =0$) results in,
\begin{equation}
    \dot{\vartheta} = \left( 1 + \gamma \left( c_{\text{tau}} \right) \right) G_0 - \frac{k_\vartheta}{\eta_\vartheta} \left( \vartheta - 1 \right) ,
\end{equation}

\noindent with the ratio $k_\vartheta/\eta_\vartheta$ establishing the characteristic atrophy relaxation time and non-negative dissipation being guaranteed - since $\Phi_a$ is convex and non-negative, the atrophy dissipation term $\partial_\vartheta \Phi_a \cdot \dot{\vartheta}$ is non-negative. The model by Sch{\"a}fer \textit{et al.} \cite{Schafer2019CMA} can be recovered with $k_\vartheta=0$ and, for simplicity, this will be the choice adopted here. However, to ensure a continuous variation of $\gamma (c_{\text{tau}} )$, we propose the following atrophy rate degradation law,
\begin{equation}
\gamma(c_{\text{tau}}) = \frac{G_c}{G_0}\,\frac{1}{1+\exp\!\left(-\kappa \left(\bar{c}_{\text{tau}}-\bar{c}_{\text{tau}}^{\text{crit}}\right)\right)}
\label{eq:gamma_smooth}.
\end{equation}

\noindent Here, $\kappa$ controls the sharpness of the transition between the two states and is set to 100 in this study. Eq. (\ref{eq:gamma_smooth}), while continuous, captures the main features of the Heaviside function-based definition of $\gamma$ by Sch{\"a}fer \textit{et al.} \cite{Schafer2019CMA}. Namely, an atrophy rate that follows the healthy ageing rate $G_0$ until a critical misfolded tau concentration is reached, after which the atrophy rate is increased by an additional contribution $G_c$, which is also a negative quantity. This behaviour is shown in Fig. \ref{fig:gammaccrit}, where the atrophy degradation function is plotted against misfolded tau concentration, as predicted by our proposed definition (\ref{eq:gamma_smooth}) and that by Sch{\"a}fer \textit{et al.} \cite{Schafer2019CMA}. We emphasise that Eq. (\ref{eq:gamma_smooth}) is defined in terms of the normalised misfolded tau concentration $\bar{c}_{\text{tau}}=c_{\text{tau}}/c_{\text{tau}}^{\text{lim}}$, to ensure consistency with Ref. \cite{Schafer2019CMA} and adopt the same value of critical concentration; using $c_{\text{tau}}$ instead would only require utilising $c_{\text{tau}}^{\text{lim}}$ (i.e., re-scaling $\bar{c}_{\text{tau}}^{\text{crit}}$).

\subsection{Summary of the governing equations}
\label{sec:Summary}

Let us conclude the formulation of our theory by summarising the governing equations of the coupled bio-chemo-mechanical problem under consideration, using their definitions in the current configuration $\mathcal{B}_t$, and discussing the existing couplings. \\

\noindent \textbf{Bio-chemical subsystem: } The bio-chemical subsystem describes the spatiotemporal evolution of toxic protein concentrations, which act as drivers of cerebral atrophy. For both proteins, we define a normalised concentration $\bar{c}_i=c_i/c_i^{\text{lim}}$, and the governing equations are written in terms of these dimensionless variables. Accordingly, and considering the protein-specific diffusivity definitions (\ref{eq:dif_ab}) and (\ref{eq:d_tauFinal}), Eqs. (\ref{eq:AbProt}) and (\ref{eq:ctau}) are reformulated as,
\begin{equation}\label{eq:abNorm}
\frac{\partial \bar{c}_{\text{A}\beta}}{\partial t}= \text{div} \left[ \left( d^{\text{ext}} \bm{i} \right)  \nabla \bar{c}_{\text{A}\beta} \right] + \alpha_{\text{A}\beta} \bar{c}_{\text{A}\beta} \left(1 - \bar{c}_{\text{A}\beta} \right),
\end{equation}
\begin{equation} \label{eq:ctauNorm}
    \frac{\partial \bar{c}_\text{tau}}{\partial t}=\text{div} \, \left[ \left(d^{\text{ext}}\bm{i}+d^{\text{axn}} \frac{\bm{F} \cdot \mathbf{a} \otimes \bm{F} \cdot \mathbf{a}}{\lambda_a^2} \right) \nabla \bar{c}_\text{tau} \right] + \alpha_\text{tau} (\bar{c}_{\text{A} \beta}) \bar{c}_\text{tau} \left(1-\bar{c}_{\text{tau}} \right).
\end{equation}

While the spread of amyloid-$\beta$ protein, Eq. (\ref{eq:abNorm}), is defined in isolation from the other bio-chemo-mechanical processes, its primary field $\bar{c}_{\text{A}\beta}$, influences tau transport through the variable $\alpha_\text{tau} (\bar{c}_{\text{A} \beta})$ - see Eq. (\ref{eq:ctauNorm}). Eq. (\ref{eq:ctauNorm}) also showcases how deformation and atrophy impact protein transport, with the deformation tensor $\bm{F}$, which includes elastic and atrophy contributions, modulating diffusion along axonal fibres. The anisotropic description of axonal tau diffusivity also incorporates the role of fibre orientation.\\

\noindent \textbf{Bio-mechanical subsystem: } The mechanical subsystem describes brain tissue deformation induced by
atrophy-driven volumetric shrinkage. The degree of atrophy is described by means of a degradation variable $\vartheta$, which is governed by a local evolution law driven by the accumulation of toxic tau protein:
\begin{equation}\label{eq:vartheta_Time}
    \dot{\vartheta} =  \left( 1 + \frac{G_c}{G_0}\,\frac{1}{1+\exp\!\left(- 10^2\left(\bar{c}_{\text{tau}}-\bar{c}_{\text{tau}}^{\text{crit}}\right)\right)} \right) G_0.
\end{equation}

The resulting volumetric shrinkage leads to tissue deformation, which is described by the atrophy-weighted quasi-static balance of linear momentum:
\begin{equation}\label{eq:finaleqmech}
    \text{div}_{\mathbf{x}} \,  \left\{ \vartheta \left[ \frac{G}{J} \left( \bm{F}^e \left( \bm{F}^{e} \right)^{T}   - \bm{I} \right) + \frac{\lambda \ln J^e}{J} \bm{I} \right] \right\} + \mathbf{B} = 0. 
\end{equation}

Eqs. (\ref{eq:abNorm}), (\ref{eq:ctauNorm}) and (\ref{eq:finaleqmech}) are discretised and solved using the finite element method with the displacement field and the concentrations of misfolded amyloid-$\beta$ and tau proteins being the degrees of freedom. In contrast, the atrophy variable $\vartheta$ is modelled as an internal variable and evolves locally in time according to the ordinary differential equation (\ref{eq:vartheta_Time}).

\section{Computational framework}
\label{Sec:Framework}

\begin{figure}[H]
    \noindent 
    \makebox[\textwidth][c]{%
        \includegraphics[width=1.2\textwidth]{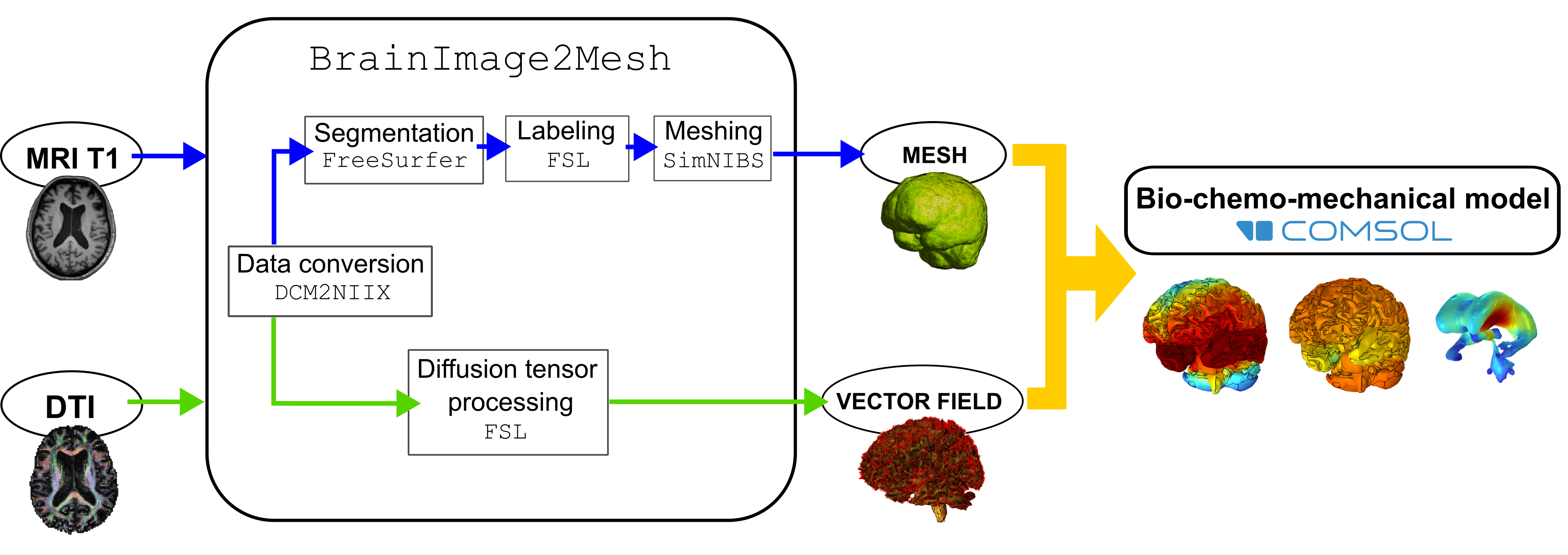} 
    }
    \caption{Schematic overview of the computational framework developed, involving two main constituents: (i) an automated processing pipeline (denoted \texttt{BrainImage2Mesh}), which takes MRI and DTI images and delivers subject-specific simulation inputs (geometry, mesh, vector field), and (ii) a bio-chemo-mechanical model, following the theory presented here, implemented in the commercial finite element package \texttt{COMSOL}.}
    \label{fig:framework}
\end{figure}

In this section, we present the computational framework introduced in this study. The framework is organised into two main components: (i) an automated medical image processing pipeline to generate subject-specific simulation inputs, and (ii) a coupled bio-chemo-mechanical finite element model to simulate protein propagation and cerebral atrophy. An overview of the workflow is shown in Fig. \ref{fig:framework}.

\subsection{Acquisition of medical images}
\label{sec:Medical image}

The first step involves acquiring medical imaging data, in the form of a structural magnetic resonance image (MRI) and a diffusion tensor imaging (DTI) scan. In this work, we acquired this data from the Alzheimer’s Disease Neuroimaging Initiative (ADNI) database \cite{ADNI}. The imaging datasets are provided in DICOM format and correspond to volumetric acquisitions defined in an orthogonal coordinate system. Each axis of this system corresponds to one anatomical orientation (sagittal, axial, or coronal) and contains a stack of approximately 200 to 250 two-dimensional slices acquired along that direction. Although the data are commonly visualised as two-dimensional sagittal, axial, and coronal slices, both MRI and DTI scans intrinsically represent three-dimensional volumetric datasets, since a single archive contains the complete spatial information of the brain. T1-weighted MRI scans are employed to capture the anatomical geometry of the brain, whereas DTI scans provide information on the preferential orientation of axonal fibres. Accordingly, T1-weighted MRI data are used for mesh generation, while DTI data are used to reconstruct the axonal orientation field.\\

\begin{figure}[h]
\centering
\begin{subfigure}{0.48\textwidth}
    \centering
    \includegraphics[width=\textwidth]{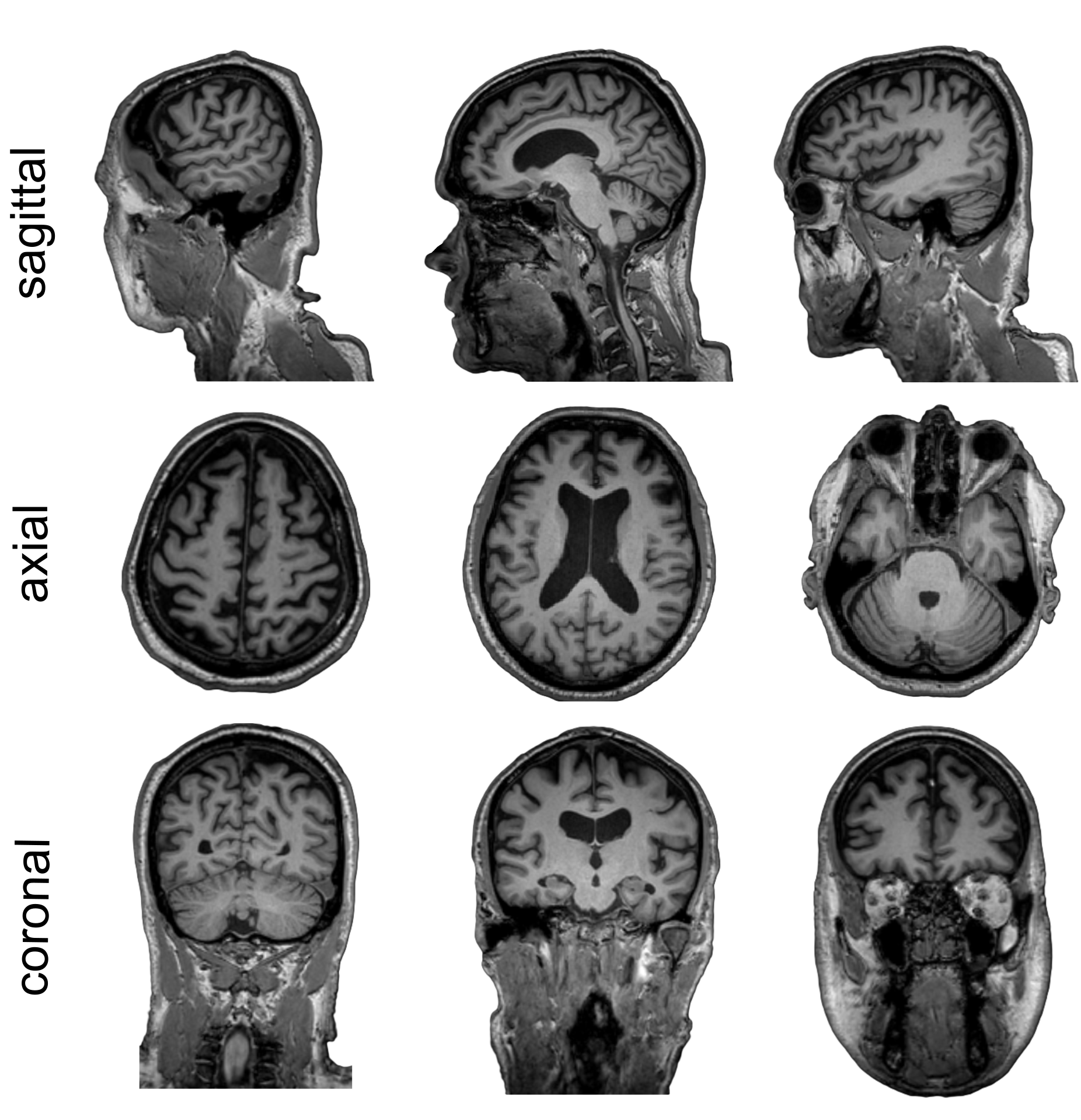}
    \caption{}
\end{subfigure}
\hfill
\begin{subfigure}{0.48\textwidth}
    \centering
    \includegraphics[width=\textwidth]{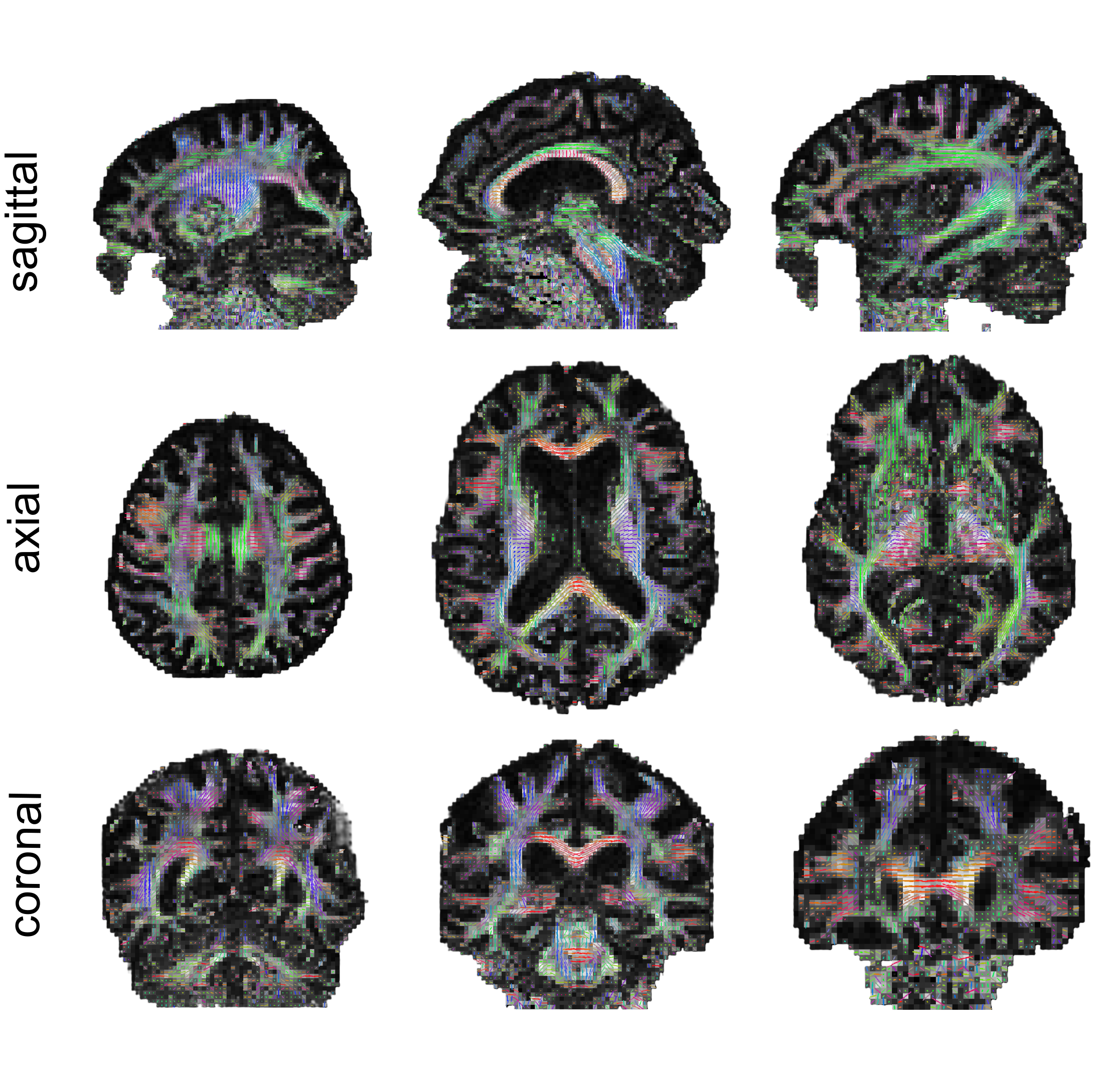}
    \caption{}
\end{subfigure}
\caption{Figure 4 shows representative subject-specific medical imaging data used in this study. Subfigure (a) displays raw T1-weighted structural MRI data acquired in DICOM format. Subfigure (b) shows post-processed diffusion tensor imaging (DTI) data, also acquired in DICOM format. In the DTI images, the diffusion tensor is estimated for each voxel. The data are visualised using a colour-coded representation in which colour indicates the orientation of the principal diffusion direction. Although visualised here as two-dimensional sagittal, axial, and coronal slices, both imaging modalities correspond to three-dimensional volumetric datasets.}
\label{fig:medical_im}
\end{figure}

Representative examples are shown in Fig.~\ref{fig:medical_im}. Specifically, we display the raw T1-weighted MRI scan and a post-processed visualisation of the DTI data. The latter is presented for visualisation purposes, where a coloured line is displayed within each voxel to indicate the local principal diffusion direction.

\subsection{Custom Linux-based preprocessing pipeline: \texttt{BrainImage2Mesh}}
\label{sec:linuxpipe}

A custom, fully automated preprocessing tool, termed \texttt{BrainImage2Mesh}, is developed in this work to generate subject-specific finite element meshes and axonal orientation fields directly from medical imaging data. The tool is provided as an executable for Linux and coordinates a series of image processing and meshing operations by interacting with multiple widely used neuroimaging software packages, all of which need to be installed beforehand. The full preprocessing pipeline requires approximately 4 hours per subject on a standard workstation.\\

The preprocessing workflow requires two input files in DICOM format: a T1-weighted MRI scan and a diffusion tensor imaging (DTI) scan. Upon execution, \texttt{BrainImage2Mesh} creates a subject-specific directory structure that organises all intermediate and final outputs of the preprocessing pipeline. Intermediate results generated by the different software packages are stored in dedicated subfolders, while the final outputs of the pipeline (mesh and axonal field) are saved in a dedicated results directory.

\subsubsection{Brain segmentation}
\label{sec:Brain seg}

As an initial step, the T1-weighted MRI DICOM image is converted into the NIfTI file format using the \texttt{dcm2nii} conversion tool \cite{dcm2niix}. The resulting NIfTI file is subsequently used for anatomical brain segmentation.\\

\begin{figure}[h]
    \centering    \includegraphics[width=0.7\textwidth]{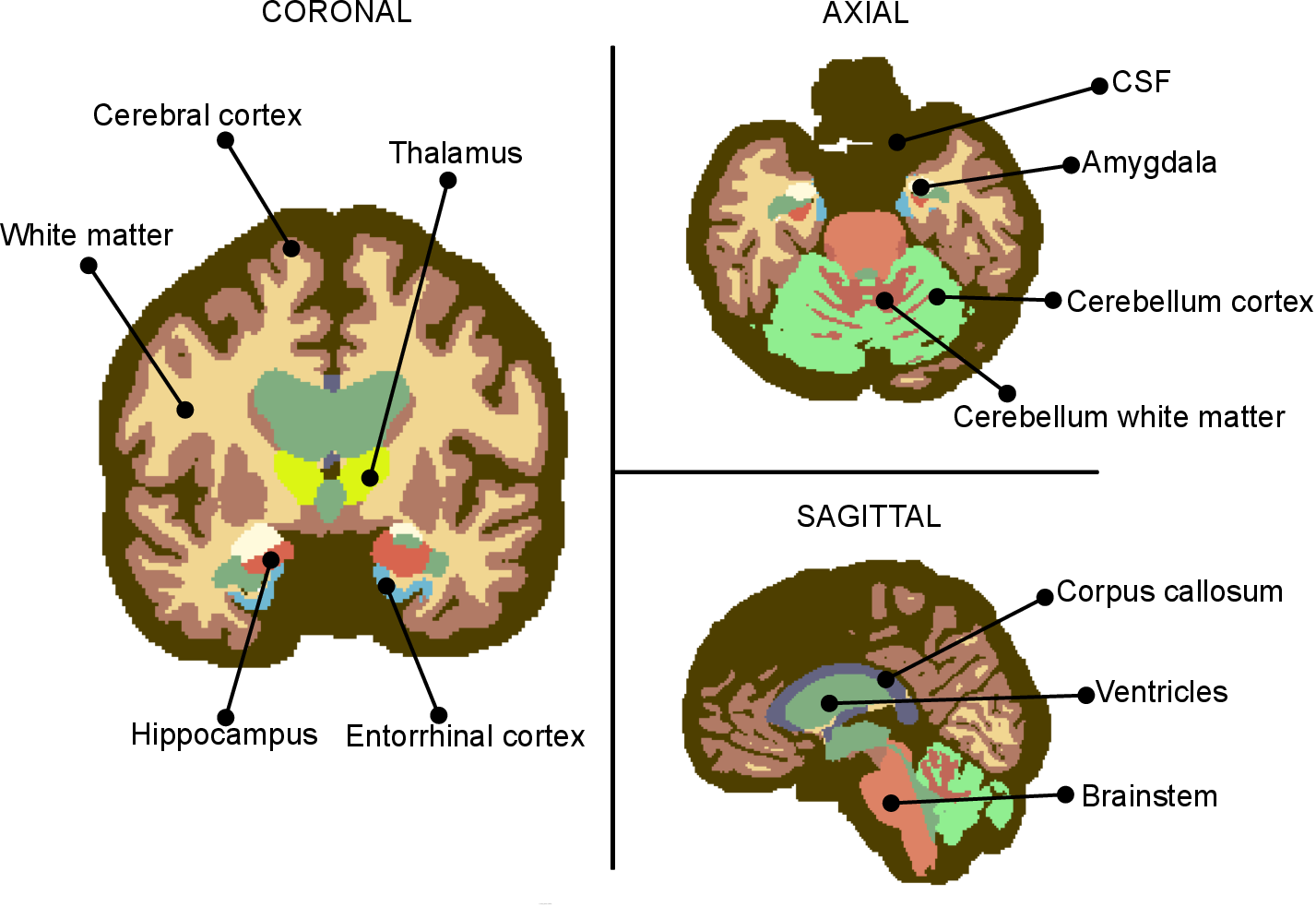}
    \caption{Brain segmentation performed on T1-weighted MRI data. The brain is partitioned into functionally relevant regions that are preserved in the computational model. In particular, selected grey and white matter subregions associated with Alzheimer’s disease pathology are retained as individual domains, while the remaining matter is grouped into homogeneous cortical and white matter regions. Ventricles and cerebrospinal fluid are treated as separate fluid domains.}
    \label{fig:seg}
\end{figure}

Brain segmentation is performed using \texttt{FreeSurfer} \cite{freesurfer}, which provides a detailed parcellation of cortical and subcortical brain structures. To reduce the overall complexity of the finite element model while preserving the anatomical regions most relevant to Alzheimer’s disease, the original \texttt{FreeSurfer} segmentation is regrouped into a reduced set of 12 anatomical domains. We distinguish between the lateral ventricles and the cerebrospinal fluid (CSF), as both are fluid-filled domains that require specific material properties, and lateral ventricular enlargement is a key hallmark of Alzheimer's disease. Within the grey matter, we preserve the hippocampus, entorhinal cortex, and amygdala, as these regions are known to act as early seed regions in Alzheimer’s disease. The remaining grey matter is grouped into a single cerebral cortex domain. For the white matter, we explicitly preserve the brainstem, cerebellar white matter, and corpus callosum, as they represent the most significant white matter regions. The remaining white matter is grouped into a homogeneous white matter domain.\\

To achieve full coverage of the brain tissue and to eliminate gaps between adjacent tissue regions, the cerebrospinal fluid (CSF) region is expanded, using morphological dilation operations implemented in \texttt{FSL} \cite{fsl}. Following the regrouping of the original \texttt{FreeSurfer} parcellation, each newly defined anatomical domain is assigned a unique integer label using \texttt{FSL}. In this way, a labelled brain volume in NIfTI format is obtained, in which each voxel is associated with a single anatomical region through its label value. An illustrative example of the labelled segmented brain is given in Fig. \ref{fig:seg}. As the same label identifiers are used for all subjects, corresponding anatomical regions share the same label, independently of differences in size, shape, or relative position. This uniform labelling enables the automated assignment of domain-specific properties in the finite element simulations.

\subsubsection{Mesh generation}
\label{sec:Mesh}

\begin{figure}[H]
    \centering
    \includegraphics[width=0.7\textwidth]{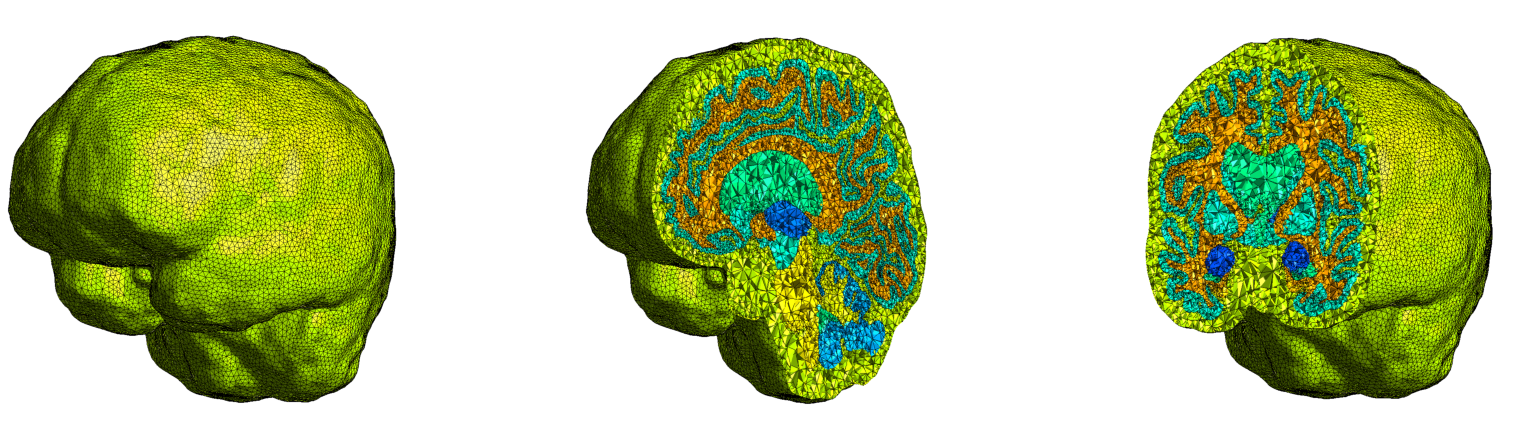}
    \caption{Volumetric tetrahedral finite element mesh generated from MRI data using the proposed pipeline. Representative sagittal (middle) and coronal (right) cross-sections are shown to illustrate internal anatomy, regional labelling, and mesh quality.}
    \label{fig:mesh}
\end{figure}

Following segmentation, \texttt{SimNIBS} \cite{simnibs} is employed to construct a volumetric finite element mesh from the updated labeled MRI volume in NIfTI format. The resulting mesh consists of first-order (linear) tetrahedral elements and explicitly preserves the 12 predefined anatomical regions as distinct subdomains, as illustrated in Fig.~\ref{fig:mesh}. The figure shows the complete tetrahedral mesh together with representative sagittal and coronal cross-sections, which are selected to highlight the internal brain anatomy and the preservation of regional boundaries. These cuts are chosen because they intersect the main anatomical structures of interest and allow visual verification of mesh quality, regional labelling, and element distribution throughout the volume. This meshing strategy enables region-specific assignment of material properties and model parameters in subsequent simulations. Using the proposed pipeline, the resulting meshes typically involve on the order of $10^7$ degrees of freedom (DOFs). The final volumetric mesh is exported in the \texttt{.bdf} file format.

\subsubsection{Vector field: axonal orientation reconstruction}
\label{sec:Vector field}

Processing of the vector field requires an initial step, in which the input DTI scan in DICOM format is converted into the NIfTI file format using \texttt{dcm2niix}. The resulting NIfTI volume is then processed using \texttt{FSL} to estimate the diffusion tensor at each voxel. From this tensor, the principal eigenvector (i.e., the eigenvector corresponding to the largest eigenvalue) is extracted, providing the local dominant diffusion direction inferred from the imaging data. The diffusion tensor $\bm{D}$ is a symmetric second-order tensor that characterises the magnitude and preferential directions of diffusion within each voxel. Diffusion tensor imaging (DTI) inherently encodes both the directionality and the magnitude of diffusion through the eigenvectors and eigenvalues of the diffusion tensor, respectively. The principal eigenvalue $\lambda_1$, associated with the principal diffusion direction, could be used to locally define the axonal diffusion magnitude. Alternatively, the mean diffusivity (MD), computed as the average of the three eigenvalues of the diffusion tensor, would provide a more robust measure of diffusion. This would allow the diffusion coefficients to vary spatially. \\

In the present formulation, the diffusion magnitudes (eigenvalues) are not taken directly from the imaging data for simplicity. Instead, the model prescribes diffusion intensity through two scalar coefficients, $d_{\mathrm{ext}}$ and $d_{\mathrm{axn}}$, which control extracellular diffusion and axonal transport. The imaging data are therefore used exclusively to provide directional information. The principal eigenvectors of the diffusion tensor are extracted to construct a vector field that specifies the local direction of diffusion, while the diffusion intensities are supplied by the model parameters. The distinction between isotropic and anisotropic behavior is applied at the tissue level, grey matter is assumed to be fully isotropic, while white matter is modeled as anisotropic. The resulting axonal orientation field assigns a unit vector to each mesh node, corresponding to the axonal orientation vector $\mathbf{a}$ introduced in the theoretical formulation and used to prescribe anisotropic diffusion in white matter. This information is stored in a text (\texttt{.txt}) file containing on the order of $2.5\times10^{5}$ entries. Each entry consists of six values: the first three specify the coordinates of the mesh node, and the remaining three define the components of the associated unit vector.\\

\begin{figure}[H]
    \centering
    \includegraphics[width=0.6\textwidth]{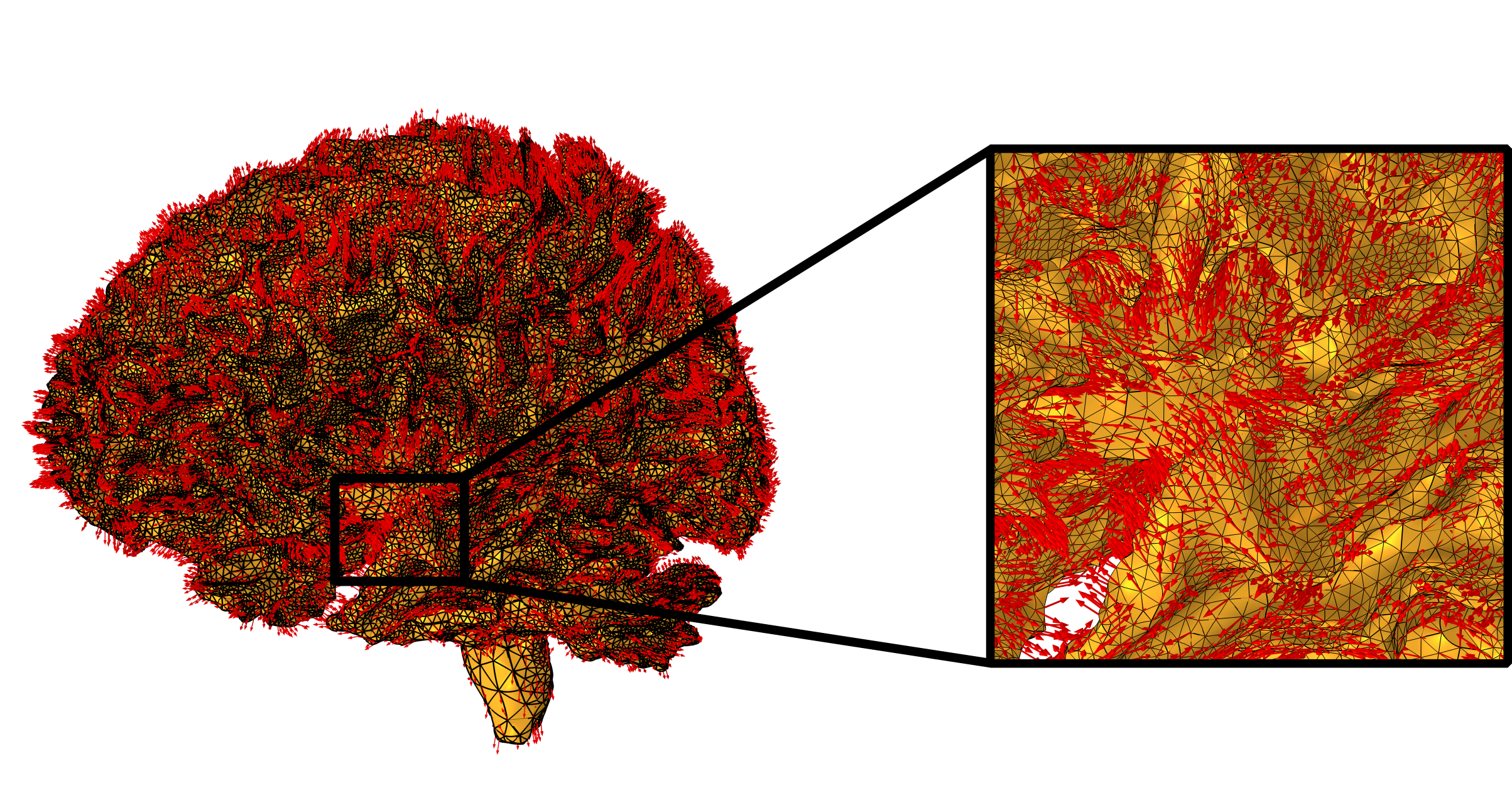}
    \caption{Axonal orientation field used to define anisotropic transport in white matter. The field is constructed from diffusion tensor imaging (DTI) data by extracting the local principal diffusion direction at each voxel and mapping it onto the finite element mesh. A unit vector representing the local dominant axonal direction is assigned at each node.}
    \label{fig:vector}
\end{figure}

\subsection{Finite element implementation and analysis}
\label{Sec:FEM}

The coupled bio-chemo-mechanical problem described in Section~\ref{Sec:Theory} is solved numerically using the finite element method. The implementation is carried out in the commercial finite element package \texttt{COMSOL}. Specifically, the built-in module \texttt{Solid Mechanics} is adapted to solve Eq. (\ref{eq:finaleqmech}), the built-in module \texttt{Transport in Solids} is exploited to implement Eqs. (\ref{eq:abNorm}) and (\ref{eq:ctauNorm}), and the \texttt{Mathematics} module is used to define the evolution of $\vartheta$, Eq. (\ref{eq:vartheta_Time}), by means of the \texttt{Domain ODEs and DAEs} interface. Additional details of the \texttt{COMSOL} implementation are provided in the Supplementary Material.\\

The boundary value problem is similar for all case studies and are characterised by the following initial and boundary conditions. For the bio-chemical problem, initial seeding regions are inferred from post-mortem neuropathological studies of Alzheimer’s disease \cite{Braak1991}. Accordingly, the initial seed concentration for the tau protein $\bar{c}_{\text{tau}_0}=0.2$ is defined in the entorhinal cortex, being zero elsewhere. For the amyloid-$\beta$ protein, the initial seed ($\bar{c}_{\text{A}\beta_0}=0.4$) is prescribed in the cerebral cortex, and is initially zero in all the other regions. In terms of boundary conditions, no transport is allowed through ventricular or cerebrospinal fluid regions (i.e., no flux Neumann boundary conditions outside brain matter). For the bio-mechanical problem, we initialise the atrophy measure to the healthy brain volume fraction ($\vartheta = 1$) and define the initial displacement field to be zero elsewhere. As boundary conditions, the displacement field is constrained ($\mathbf{u}=\mathbf{0}$) in the external surface of the cerebrospinal fluid domain to mimic the mechanical constraint imposed by the skull. 

\section{Results}
\label{Sec:Results}

\begin{table}[h]
\centering
\caption{Elastic properties and atrophy rates assigned to the different brain regions, in agreement with the related literature \cite{Blinkouskaya2021Frontiers,Schafer2019CMA}.}
\label{tab:material_parameters}
\begin{tabular}{l l c c c c}
\hline
\textbf{Type} & \textbf{Region} & ${\lambda}$ [kPa] & ${G}$ [kPa] & $G_c$ [1/year] & $G_0$ [1/year] \\
\hline
\multirow{2}{*}{CSF liquid} 
& Ventricles & 1.66 & 0.50 & -- & -- \\
& CSF & 7.22 & 14.43 & -- & -- \\
\hline
\multirow{6}{*}{Grey matter}
& Amygdala & 32.33 & 1.0 & $-6.0\times10^{-3}$ & $-6.0\times10^{-3}$ \\
& Hippocampus & 32.33 & 1.0 & $-6.0\times10^{-3}$ & $-6.0\times10^{-3}$ \\
& Thalamus & 32.33 & 1.0 & $-6.0\times10^{-3}$ & $-6.0\times10^{-3}$ \\
& Entorhinal cortex & 32.33 & 1.0 & $-7.0\times10^{-3}$ & $-7.0\times10^{-3}$ \\
& Cerebral cortex & 32.33 & 1.0 & $-6.0\times10^{-3}$ & $-6.0\times10^{-3}$ \\
& Cerebellum cortex & 32.33 & 1.0 & $-6.0\times10^{-3}$ & $-6.0\times10^{-3}$ \\
\hline
\multirow{4}{*}{White matter}
& Cerebellum WM & 64.67 & 2.0 & $-5.5\times10^{-3}$ & $-5.5\times10^{-3}$ \\
& Brainstem & 64.67 & 2.0 & $-5.5\times10^{-3}$ & $-5.5\times10^{-3}$ \\
& White matter & 64.67 & 2.0 & $-5.5\times10^{-3}$ & $-5.5\times10^{-3}$ \\
& Corpus callosum & 64.67 & 0.7 & $-5.5\times10^{-3}$ & $-5.5\times10^{-3}$ \\
\hline
\end{tabular}
\end{table}

\begin{table}[h]
\centering
\caption{Biophysical parameters adopted in the model, consistent with the literature \cite{Blinkouskaya2021Frontiers,Schafer2019CMA,antonietti2025numerical,Pal2022SciRep}.}
\label{tab:biophysical_parameters}
\setlength{\tabcolsep}{6pt}
\renewcommand{\arraystretch}{1.15}

\begin{tabular}{cccccccc}
\hline
$d_{\mathrm{ext}}$ &
$d_{\mathrm{axn}}$ &
$\bar{c}^{\mathrm{crit}}_{\text{tau}}$ &
$\bar{c}_{\text{tau}}^0$ &
$\bar{c}_{\mathrm{A}\beta}^0$ &
$\alpha_{\text{tau}}$ &
$\alpha_{\mathrm{A}\beta}$ &
$\tilde{a}_0$ \\[-0.6ex]
{[mm$^2$/year]} &
{[mm$^2$/year]} &
{[-]} &
{[-]} &
{[-]} &
{[1/year]} &
{[1/year]} &
{[1/year]} \\
\hline
8.0 &
80.0 &
0.2 &
0.4 &
0.2 &
$1.09(0.4\bar{c}_{\text{tau}}+1)^{-0.55}$ &
0.2 &
1.04 \\
\hline
\end{tabular}

\vspace{2mm}

\begin{tabular}{cccccccc}
\hline
$\tilde{a}_1$ &
$\tilde{a}_{12}$ &
${a}_1$ &
$\tilde{k}_0$ &
$\tilde{k}_1$ &
$\tilde{k}_{12}$ &
${k}_1$ &
$\tilde{k}_3$ \\[-0.6ex]
{[1/year]} &
{[1/year]} &
{[1/year]} &
{[1/year]} &
{[1/year]} &
{[1/year]} &
{[1/year]} &
{[1/year]} \\
\hline
1.38 & 1.38 & 0.83 & 0.60 & 0.55 & 1.00 & 0.55 & 2.00 \\
\hline
\end{tabular}
\end{table}

We proceed to present simulation results for the brain model. All results are obtained from subject-specific finite element meshes generated following the preprocessing pipeline described in Section~\ref{sec:linuxpipe}. Imaging data from a 75-year-old male subject is adopted. Three numerical simulations are performed: an Alzheimer’s disease (AD) scenario accounting for anisotropic transport in white matter, a healthy ageing scenario considered as a reference case, and an additional AD simulation assuming isotropic transport in white matter, in order to assess the effect of neglecting anisotropy. The main focus is on the spatial distribution of toxic proteins, the associated atrophy measure, and the resulting changes in brain volume. All simulations are performed over a time horizon of 20 years to capture the long-term progression of toxic protein accumulation and the associated atrophy.
\\

The material properties assigned to the different brain regions are summarised in Table \ref{tab:material_parameters} and are consistent with values reported in the literature  \cite{Blinkouskaya2021Frontiers,Schafer2019CMA}.
In particular, we prescribe a higher atrophy rate in grey matter than in white matter, in agreement with neuroimaging studies \cite{Jack2010}.
The Poisson's ratio for brain parenchyma is defined to be close to incompressibility, with values of approximately $\nu \approx 0.48$, which we consider appropriate for modelling soft brain tissue \cite{Budday2017Acta}. In contrast, we model the ventricular region as ultrasoft and moderately compressible, with a Poisson's ratio of $\nu = 0.38$ and low shear moduli, in order to mimic fluid mechanical behavior and allow volumetric expansion. A more realistic description of ventricular mechanics could be achieved by incorporating fluid behaviour through fluid-structure interaction models, as it is, in reality, a cavity filled with cerebrospinal fluid \cite{lee2024impact}. Additionally, we assign a reduced shear modulus to the corpus callosum to reflect its comparatively soft mechanical response \cite{Fjell2014CerebralCortex} and increase the accelerated atrophy rate parameter $G_c$ in the entorhinal cortex, as this region is known to undergo pronounced atrophy \cite{Budday2017Acta}. It is worth noting that, although region-dependent variations are partially accounted for, the model still relies on homogeneous material properties within each region. Incorporating more detailed region-specific parameters could further improve the model \cite{Griffiths2023}.\\

The biophysiological parameters are summarised in Table \ref{tab:biophysical_parameters}. These parameters are inherently difficult to measure directly in vivo and are therefore commonly calibrated based on simulation outcomes. The diffusion parameters $\bm{D}_{\mathrm{axn}}$ and $\bm{D}_{\mathrm{ext}}$ were taken from values reported in the literature \cite{Schafer2019CMA,corti2023discontinuous}. For example, in the work of Schäfer et al. \cite{Schafer2019CMA}, these coefficients were selected and adjusted by comparing the resulting simulated brain volume loss with clinical imaging data. The kinetic parameters governing the propagation of tau and amyloid-$\beta$ were selected within ranges reported in the literature \cite{corti2023discontinuous}. However, a calibration step was necessary because most previous models consider tau propagation alone. In such cases, higher tau kinetic rates are typically used. Since our model incorporates the accelerating influence of amyloid-$\beta$ on tau propagation, the tau kinetic parameters were slightly reduced to maintain realistic propagation dynamics consistent with published studies.\\

The initial concentrations used as pathological seeds were also chosen based on values commonly adopted in previous computational studies \cite{Schafer2019CMA,Blinkouskaya2021Frontiers,Corti2024BrainMultiphysics}. In practice, these seeds are modelling assumptions, since the exact initial concentrations vary between subjects. We prescribe an initial toxic tau concentration seed of $\bar{c}^{\text{tau}}_0 = 0.4$ and a lower initial amyloid-$\beta$ seed of $\bar{c}^{\text{A}\beta}_0 = 0.2$, as amyloid-$\beta$ deposition is known to be more spatially diffuse and tau pathology is characterised by a concentrated accumulation \cite{Braak1991}.

\subsection{Spatiotemporal progression of toxic proteins in Alzheimer’s disease}
\label{sec:Toxic protein}

\begin{figure}[h]
    \centering
    \includegraphics[width=1\textwidth]{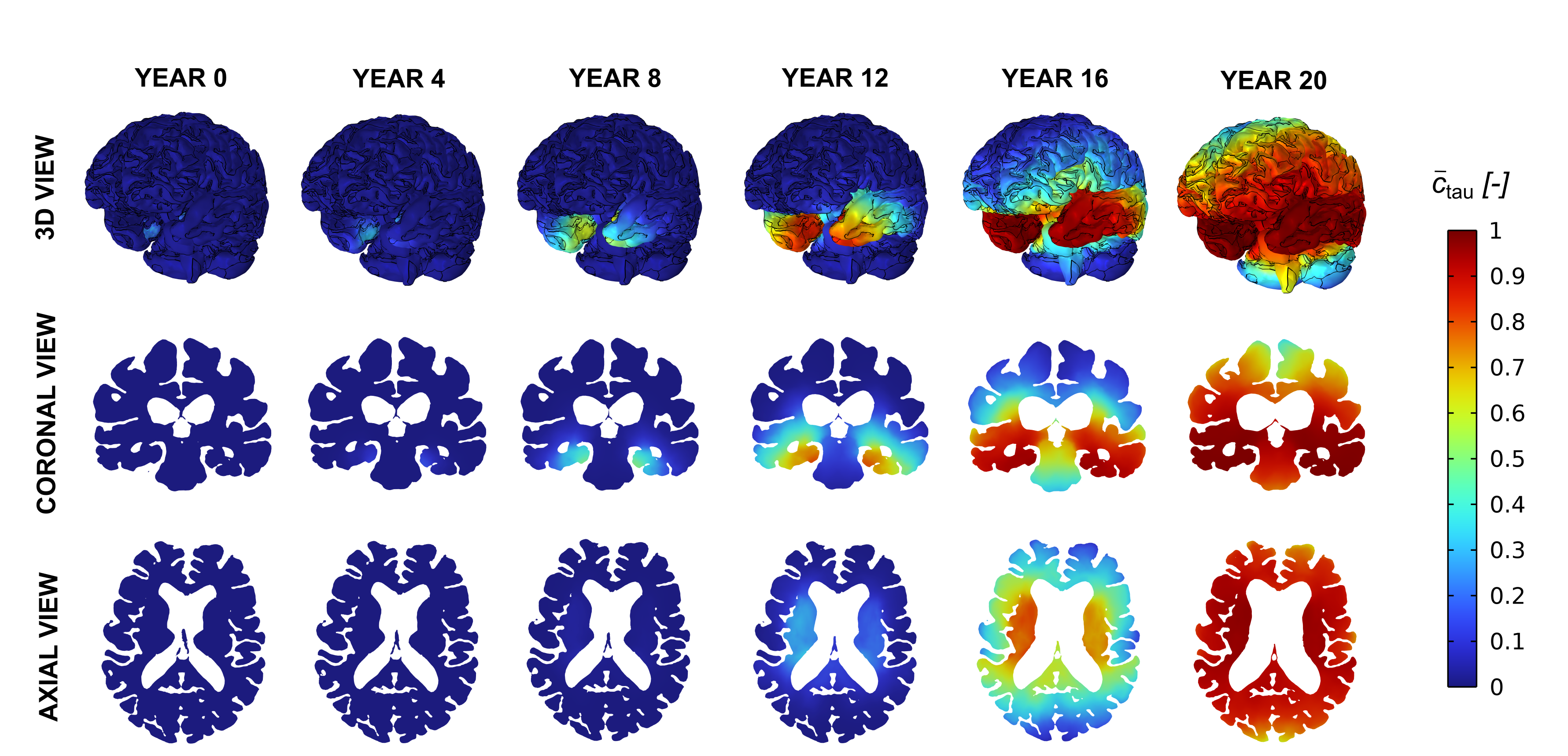}
    \caption{Spatiotemporal evolution of the normalised toxic tau protein concentration ($\bar{c}_{\text{tau}}$) over a 20-year simulation period, starting from an initial seed in the entorhinal cortex.}
    \label{fig:tau}
\end{figure}

\begin{figure}[H]
    \centering
    \includegraphics[width=1\textwidth]{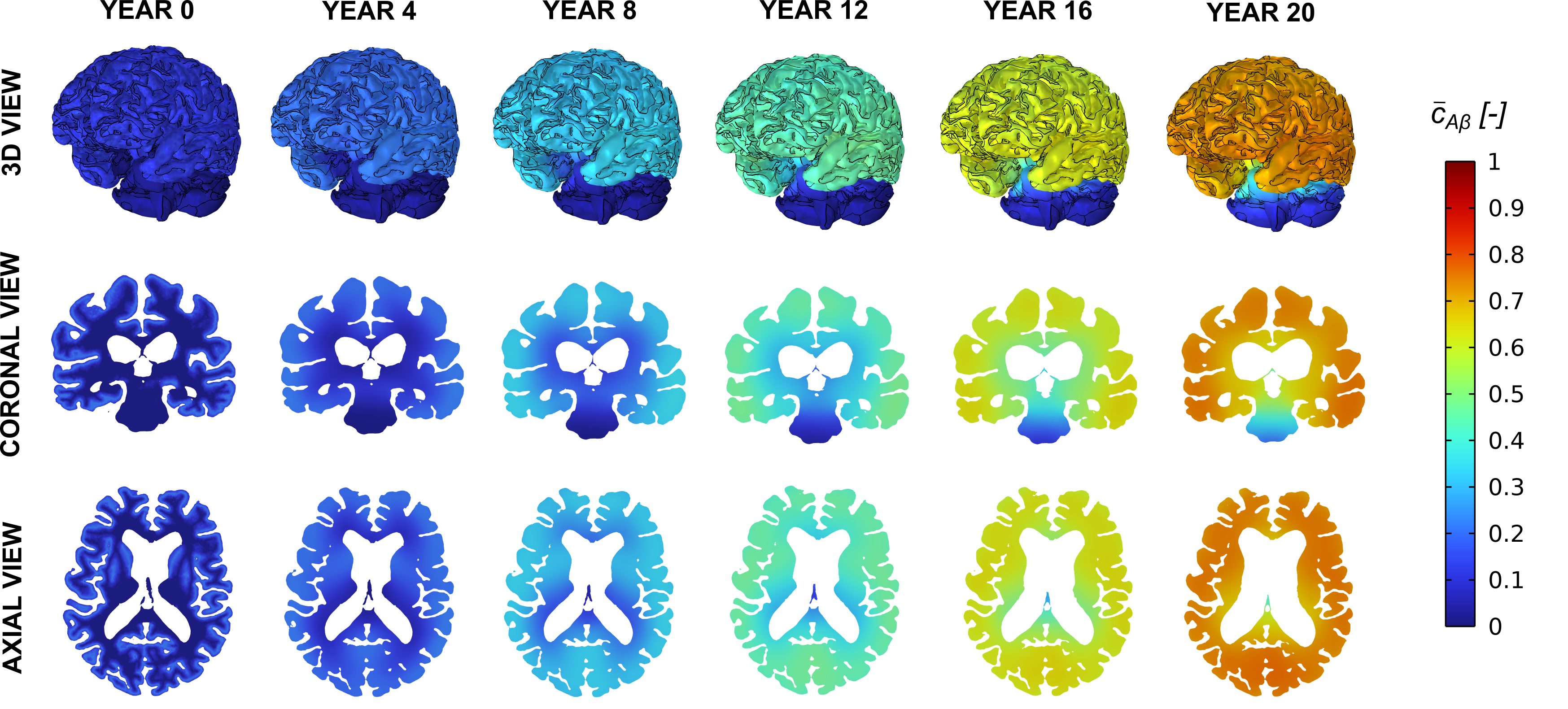}
    \caption{Spatiotemporal evolution of the normalised toxic amyloid-A$\beta$ protein concentration ($\bar{c}_{\text{A}\beta}$) over a 20-year simulation period, starting from an initial seed in the cerebral cortex.}
    \label{fig:ab}
\end{figure}

The spatial progressions of toxic tau and toxic amyloid-A$\beta$ concentrations are respectively shown in Figs. \ref{fig:tau} and \ref{fig:ab}. At early stages, toxic tau remains highly localised within the entorhinal cortex, where the initial seed is prescribed. This localised distribution persists over several years, and no significant spatial propagation is observed during the initial phase of the simulation. Then, after 12 years, a marked spatial expansion of toxic tau is observed. This delayed but rapid spreading coincides with a sufficiently widespread accumulation of amyloid-$\beta$ protein. As described by Eq.(\ref{eq:alpha_tau}), the effective growth rate of tau, $\alpha_{\text{tau}}$, increases with the local concentration of amyloid-$\beta$. This directly amplifies the reaction term in the Fisher-Kolmogorov formulation, leading to an enhanced local growth of the toxic tau concentration, $c_{\tau}$. As a result, tau propagation exhibits a sudden increase in its spatial extent once amyloid-beta reaches a critical level. In contrast, amyloid-$\beta$ displays a smoother and more gradual spatial progression throughout the simulation. Since amyloid-beta transport is modelled exclusively through extracellular diffusion and does not include any accelerating terms in the present formulation, its spatial distribution evolves more uniformly over time.

\subsection{Spatiotemporal distribution of the atrophy factor in healthy brain aging and Alzheimer’s disease}
\label{sec:Atrophy}

\begin{figure}[h]
    \centering
    \includegraphics[width=1\textwidth]{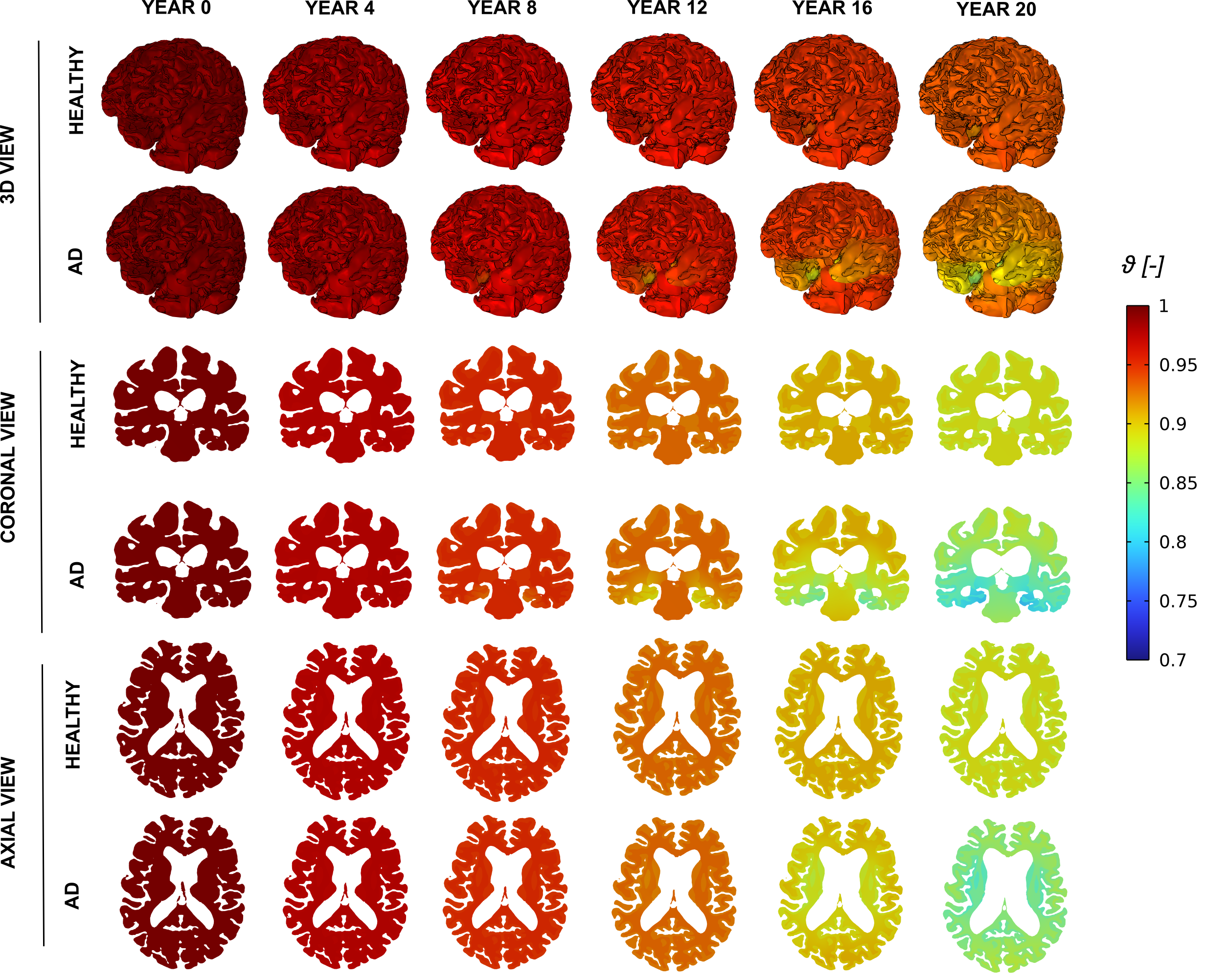}
    \caption{Spatiotemporal evolution of the cerebral atrophy measure $\vartheta$ ($1 \to 0$) in a reference healthy brain and in a brain with Alzheimer's disease (AD) over a 20-year simulation period. An atrophy measure $\vartheta=1$ indicates that 100\% of the initial brain volume remains unaffected by atrophy.}
    \label{fig:atrophy}
\end{figure}

The evolution of cerebral atrophy over time is shown in Fig. \ref{fig:atrophy}, as described by the evolution of the atrophy parameter $\vartheta$. In addition to the case of Alzheimer's disease (labelled AD), results are also shown for the case of a healthy brain, where $G_c=0$ and no protein kinetics are considered (i.e., only the bio-mechanical problem is solved). Both agree at the beginning of the simulation (year 0), where $\vartheta=1$ everywhere, indicating that 100\% of the initial brain volume remains unaffected by atrophy, but noticeable differences are seen after year 12. While the atrophy levels remain low in the healthy case, with values of the atrophy parameter generally above 0.9 even after 20 years, some regions of the Alzheimer's disease brain reach atrophy values as low as 0.7. In terms of the spatial distribution, when Alzheimer’s disease is present, the spatial distribution of atrophy correlates with the regions where higher concentrations of toxic tau were observed in Fig. \ref{fig:tau}. As a result, atrophy is more pronounced around the lateral ventricular walls and in deep brain regions such as the entorhinal cortex and the hippocampus. In contrast, the healthy reference case does not exhibit any localised regions of pronounced atrophy. In both cases, grey matter exhibits a slightly faster progression of atrophy compared to white matter, consistent with the atrophy rates prescribed in the model.

\subsection{Brain deformations in healthy brain ageing and Alzheimer’s disease}
\label{sec:Deformation}

\begin{figure}[h]
    \centering
    \includegraphics[width=0.9\textwidth]{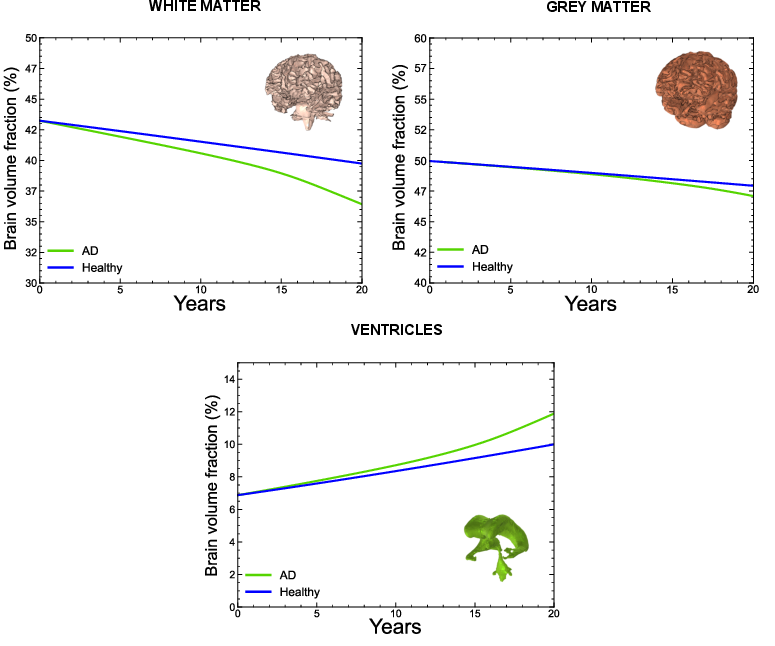}
    \caption{Temporal evolution of regional brain volumes over a 20-year simulation period for white matter, grey matter, and ventricles. Volumes are expressed as percentages relative to the total initial brain volume at $t=0$. Results are shown for both the AD case and the healthy reference case.}
    \label{fig:brainvolume}
\end{figure}

To quantify brain deformation, we monitor the temporal evolution of regional brain volumes throughout the simulation for both the Alzheimer’s disease (AD) and the healthy reference case. For simplification of the analysis, we group anatomical brain matter regions into grey matter and white matter categories, as defined in Table \ref{tab:material_parameters}. In addition to brain matter, we also consider the ventricles, as their enlargement is a key morphological hallmark of Alzheimer’s disease.\\

We express the evolution of grey matter, white matter, and ventricular volumes as percentages relative to the total initial volume. We obtained this reference volume at the initial time by summing the corresponding grey matter, white matter, and ventricular volumes. We did not include cerebrospinal fluid as it was artificially expanded as described in Section \ref{sec:linuxpipe} and it is not the focus of the present study. Our model predicts that grey matter volume decreases from 42.8\% to 39.7\% in the healthy case and to 36.2\% in the presence of Alzheimer's disease. White matter volume drops from 50.0\% to 47.6\% in the healthy case and to 46.7\% in the Alzheimer's disease case. Ventricular volume increases from 6.8\% to 10.0\% in the healthy case and to 11.9\% in the Alzheimer's disease case. This volumetric fraction evolution is shown in Fig. \ref{fig:brainvolume}, where it can be observed that the healthy case exhibits a linear evolution, whereas the Alzheimer's disease case shows a nonlinear progression.\\

\begin{figure}[h]
    \centering
    \includegraphics[width=1\textwidth]{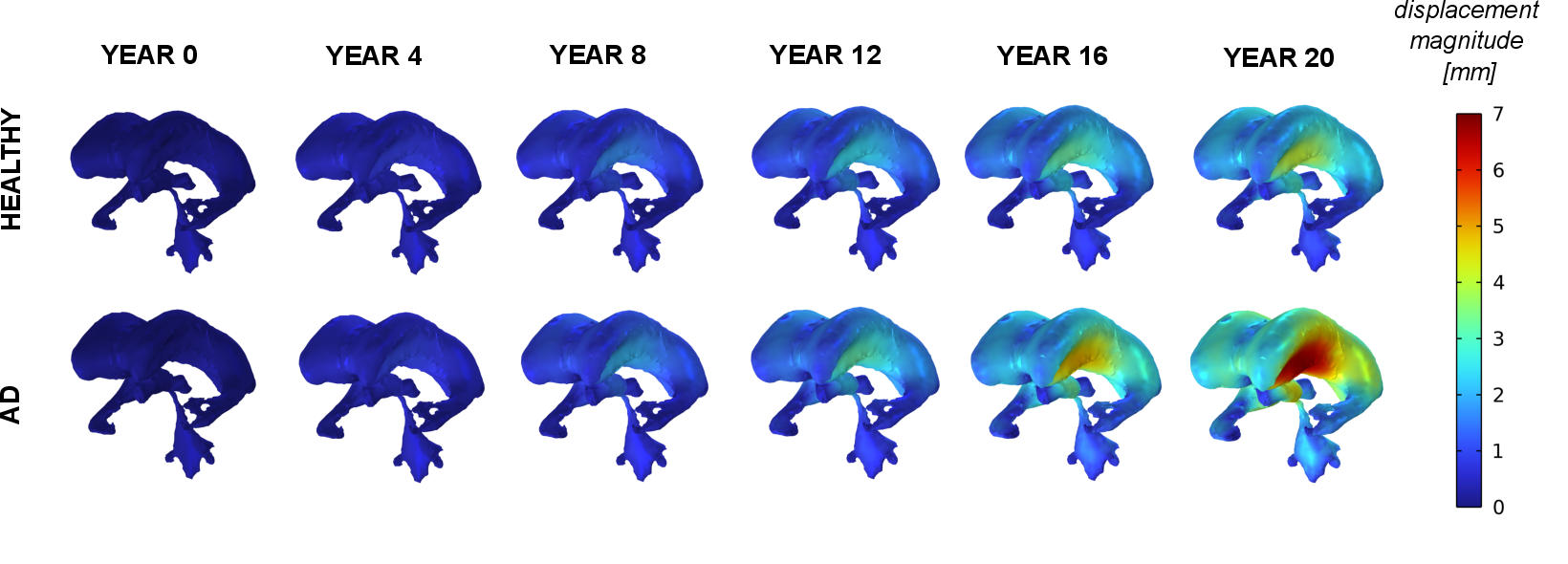}
    \caption{Spatiotemporal evolution of ventricles displacement magnitude over a 20 year simulation period for the healthy case and the AD case.}
    \label{fig:ventridisp}
\end{figure}

In addition, we illustrate in Fig. \ref{fig:ventridisp} the spatiotemporal evolution of the displacement field associated with ventricular expansion, as it represents a key morphological feature of Alzheimer’s disease. The largest deformations are concentrated along the lateral ventricular walls, particularly in the ventricular body and adjacent horns, whereas the ventricular tail exhibits smaller displacements. These findings are consistent with previous studies \cite{Apostolova2012}.\\

To evaluate whether the trends observed for a single geometry persist across a sufficiently large cohort, we repeated the same analysis for 20 subjects, 10 healthy control subjects and 10 subjects with Alzheimer’s disease, all aged between 74 and 76 years and with a male to female ratio of 1:1. Subject-specific meshes and axonal orientation fields were generated for each case, and corresponding healthy or Alzheimer’s disease simulations were performed.\\ 

\begin{figure}[h]
    \centering
    \includegraphics[width=1\textwidth]{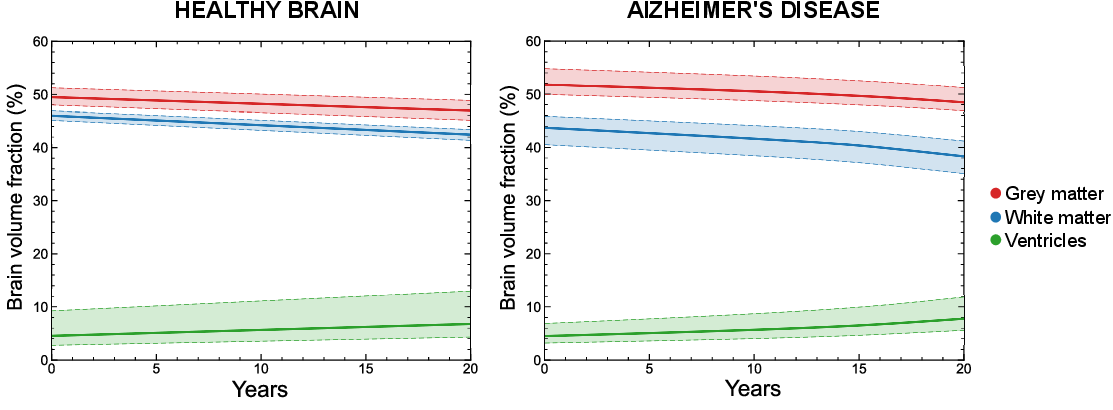}
    \caption{Temporal evolution of regional brain volumes for a cohort of 10 healthy controls (left) and 10 subjects with Alzheimer’s disease (right). Solid lines represent the cohort-averaged evolution, while shaded areas indicate subject variability. Volumes are expressed as percentages relative to the initial brain volume at $t=0$.}
    \label{fig:cohort_volumes}
\end{figure}

The results are shown in Fig. \ref{fig:cohort_volumes}, focusing on brain volume changes. Across the cohort, consistent deformation patterns are observed, although with a degree of inter-subject variability. In the healthy cohort, volume changes remain relatively moderate and approximately linear, consistent with physiological ageing \cite{Whitwell2001}. In contrast, subjects with Alzheimer’s disease show an accelerated reduction of both grey and white matter volumes, accompanied by an increase in ventricular volume. Notably, ventricular volumes exhibit a great degree of inter-subject variability, indicating sensitivity to baseline anatomy and disease progression. Specific subject analysis, provided in the Supplementary Material, revealed that subjects with larger initial ventricular volumes in both cases tend to exhibit a higher relative increase in ventricular volume over time. \\

\subsection{Effect of axonal orientation: comparison between isotropic and anisotropic models}
\label{sec:Axon}

A hypothesis and novelty of this formulation is the consideration of directional, anisotropic protein transport along the axonal orientation. To assess its importance, we compare isotropic- and anisotropic-based predictions of the spatiotemporal distribution of the misfolded tau protein concentration in white matter, which serves as a proxy for disease progression. To this end, we performed an additional Alzheimer’s disease simulation in which the axonal vector field was not considered. Instead, axonal transport was modelled as isotropic diffusion. In this case, the diffusion tensor reduces to
\begin{equation}
\bm{d} = d_{\mathrm{ext}}\,\bm{I} + \frac{d_{\mathrm{axn}}}{3}\,\bm{I},
\end{equation}
This formulation ensures that the total diffusivity remains comparable to the anisotropic case, preserving the trace of the diffusion tensor and thus the mean diffusivity. As a result, differences between isotropic and anisotropic simulations reflect directional effects rather than changes in overall diffusion magnitude.\\

\begin{figure}[H]
    \centering
    \begin{subfigure}{\textwidth}
        \centering
        \includegraphics[width=1\textwidth]{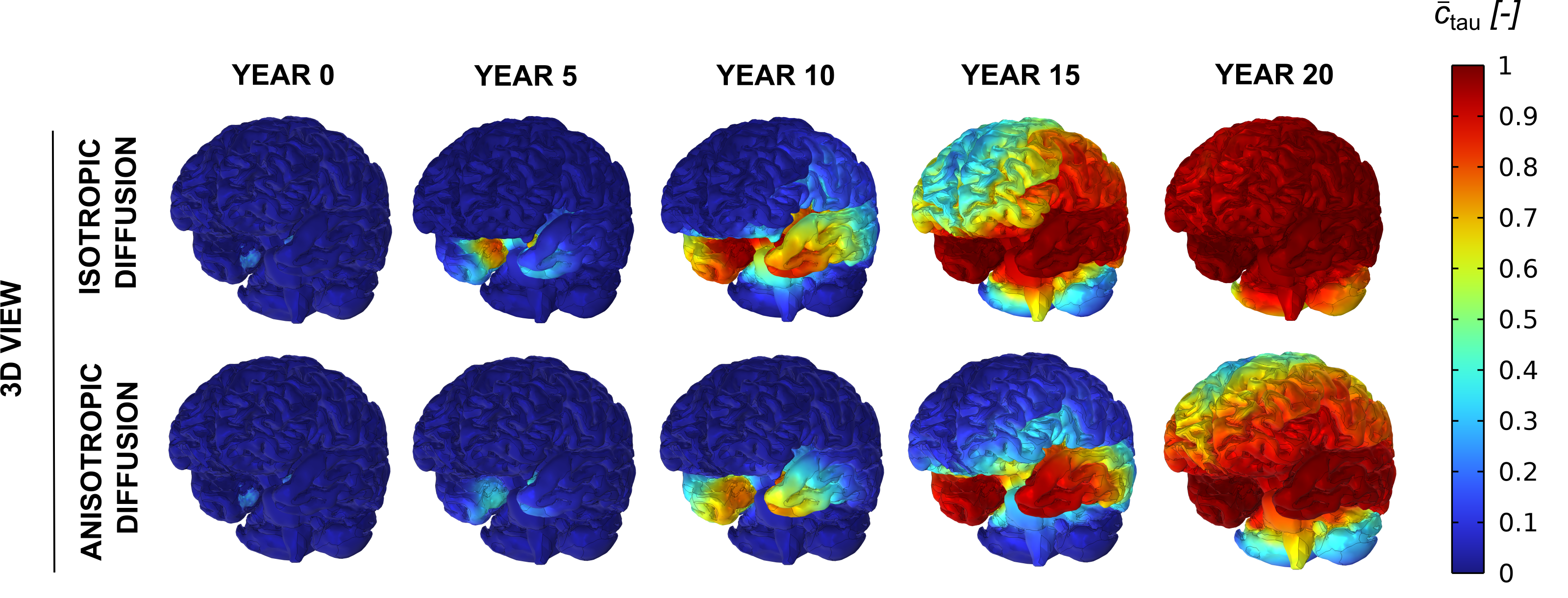}
        \caption{}
    \end{subfigure}

    \vspace{0.7cm}

    \begin{subfigure}{\textwidth}
        \centering
        \includegraphics[width=0.9\textwidth]{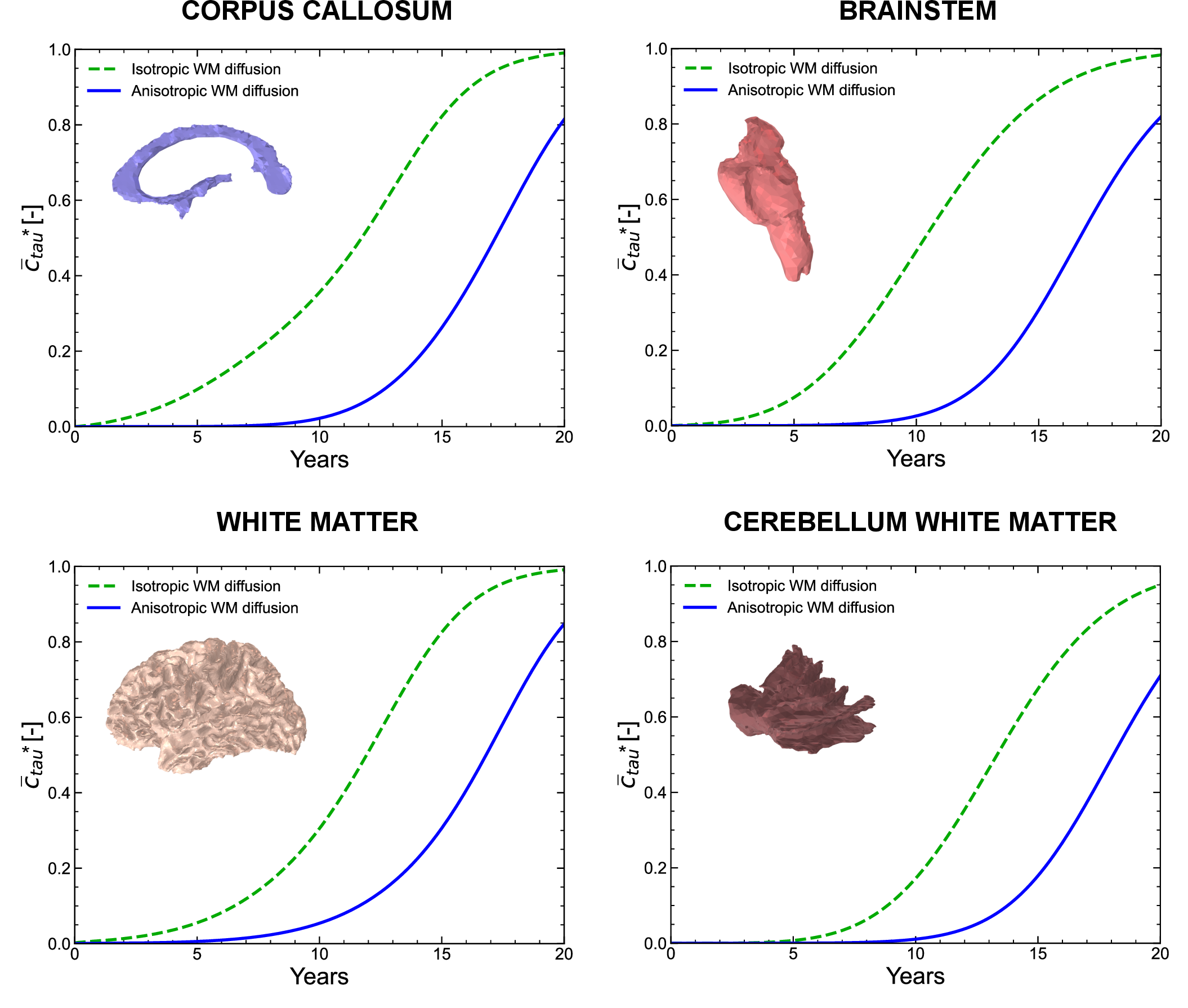}
        \caption{}
    \end{subfigure}

    \caption{Effect of axonal orientation on toxic tau propagation over a 20-year simulation. 
    (a) 3D views of toxic tau concentration obtained using isotropic and anisotropic diffusion in white matter. 
    (b) Temporal evolution of the spatially averaged toxic tau concentration for isotropic and anisotropic diffusion in regional brain volumes.}
    \label{fig:anisotropy_vs_isotropy}
\end{figure}

The results of the simulations performed with isotropic and anisotropic white matter diffusion are presented in Fig. \ref{fig:anisotropy_vs_isotropy}. The top rows show 3D views of the toxic tau normalised concentration field at representative time points, allowing visual comparison of the spatial spreading patterns obtained with and without axonal anisotropy. The bottom panels report the temporal evolution of the spatially averaged toxic tau concentration computed over the white matter regions ${\bar{c}^*_\text{tau}}$, which is used to quantify differences in overall accumulation dynamics between isotropic and anisotropic diffusion models. The results show that neglecting the axonal vector field leads to an overestimation of toxic tau concentration over time, suggesting that anisotropic diffusion along axonal pathways acts as a constraining mechanism consistent with the slow progression of tau pathology observed clinically \cite{Braak1991}. Beyond global concentration levels, clear differences also emerge in the spatial distribution of tau. The anisotropic model exhibits preferential spreading of tau along specific pathways, which is absent in the isotropic case, where the spatial distribution is more homogeneous.

\subsection{Clinical benchmarking}
\label{sec:Method}

\begin{table}[H]
\centering
\caption{Subject-specific initial conditions for toxic tau and A$\beta$
concentrations at age 75 used to initialise the Alzheimer’s disease
simulations. The table reports the seeding regions and corresponding
normalised concentration values for each subject.}
\label{tab:initial_conditions_ad}
\begin{tabular}{c l c l c}
\hline
\textbf{Subject} &
\multicolumn{2}{c}{${\bar{c}_{\text{tau}_0}}$[-]} &
\multicolumn{2}{c}{$\bar{c}_{\text{A}\beta_0}$[-]}\\
 & Seeding region & Value & Seeding region & Value \\
\hline
\multirow[c]{4}{*}{D}
 & Entorhinal cortex & 0.40 & Cerebral cortex & 0.40 \\
 & Amygdala          & 0.92 & Amygdala        & 0.30 \\
 & Hippocampus       & 0.485 & Hippocampus     & 0.14 \\
 & Thalamus          & 0.21 & Thalamus        & 0.60 \\
\hline
\multirow[c]{4}{*}{C}
 & Entorhinal cortex & 0.51 & Cerebral cortex & 0.80 \\
 & Amygdala          & 0.70 & Brainstem       & 0.50 \\
 & Hippocampus       & 0.41 & Cerebellum      & 0.60 \\
 & Thalamus          & 0.24 & Thalamus        & 0.50 \\
\hline
\end{tabular}
\end{table}

Finally, we develop a benchmarking strategy to assess the ability of the proposed framework to reproduce subject-specific volumetric changes observed in longitudinal imaging data. Separate procedures are designed for the healthy case and for the Alzheimer’s disease case, as described below.\\

\noindent \textbf{Healthy case benchmarking}. For the healthy case simulations, we selected two cognitively healthy subjects (Subject A, female, and Subject B, male) for whom structural brain images were available at ages 75 and 85 years. We generated subject-specific meshes from the images acquired at age 75, and we performed healthy simulations over a temporal horizon of 10 years. We compared the simulated brain volumes obtained at the end of the 10-year period with the corresponding volumetric measurements derived from the images acquired at age 85. Since healthy ageing in the proposed model is governed solely by a linear atrophy process, we did not require any subject-specific personalisation of the initial conditions for this validation.\\

\noindent \textbf{Alzheimer’s disease case benchmarking}. Clinical benchmarking of the Alzheimer’s disease simulations required a more elaborate procedure due to the heterogeneous nature of disease progression. We selected two subjects diagnosed with Alzheimer’s disease (Subject C, female and Subject D, male) for whom both imaging and clinical data were available at ages 75 and 80 years. The relatively short temporal interval was dictated by data scarcity. For each subject, we generated subject-specific meshes from the images acquired at age 75, and we performed simulations over a temporal horizon of 5 years. We compared the simulated brain volumes obtained at the end of the simulation with the volumetric measurements derived from the images acquired at age 80. As each subject may be at a different stage of Alzheimer’s disease and exhibit distinct initial pathological conditions, direct comparison requires subject-specific initial conditions. To this end, we extracted subject-specific biomarkers for tau and amyloid-$\beta$ from the ADNI database. Specifically, we used regional standardised uptake value ratios (SUVRs), which compare the imaging signal measured in each brain region to a reference region, providing an estimate of how much protein is present locally. Higher SUVR values indicate a larger local accumulation of the targeted protein. The SUVR values were normalised to obtain dimensionless quantities compatible with the model formulation. These normalised biomarker levels were then incorporated as initial conditions for tau and A$\beta$ in the simulations. The specific values adopted for each subject are reported in Table~\ref{tab:initial_conditions_ad}.

\begin{figure}[H]
    \centering
    \includegraphics[width=1\textwidth]{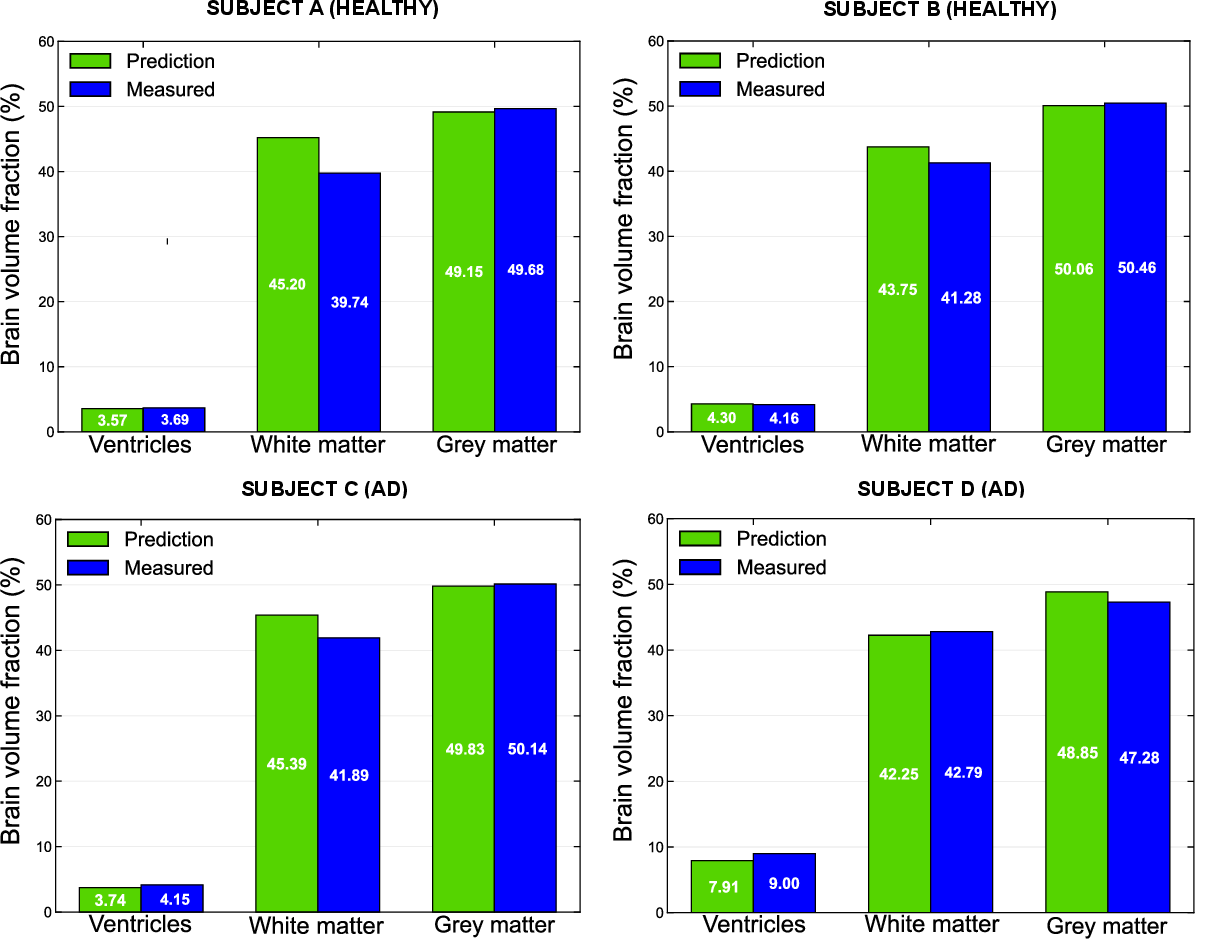}
    \caption{Comparison between predicted and measured brain volume fractions for ventricles, white matter, and grey matter. Results are shown for two healthy subjects and two subjects with Alzheimer’s disease. Predicted values correspond to the final time point of the simulations, while measured values are derived from medical imaging.}
    \label{fig:validation_volumes}
\end{figure}

\begin{figure}[h]
    \centering
    \includegraphics[width=1\textwidth]{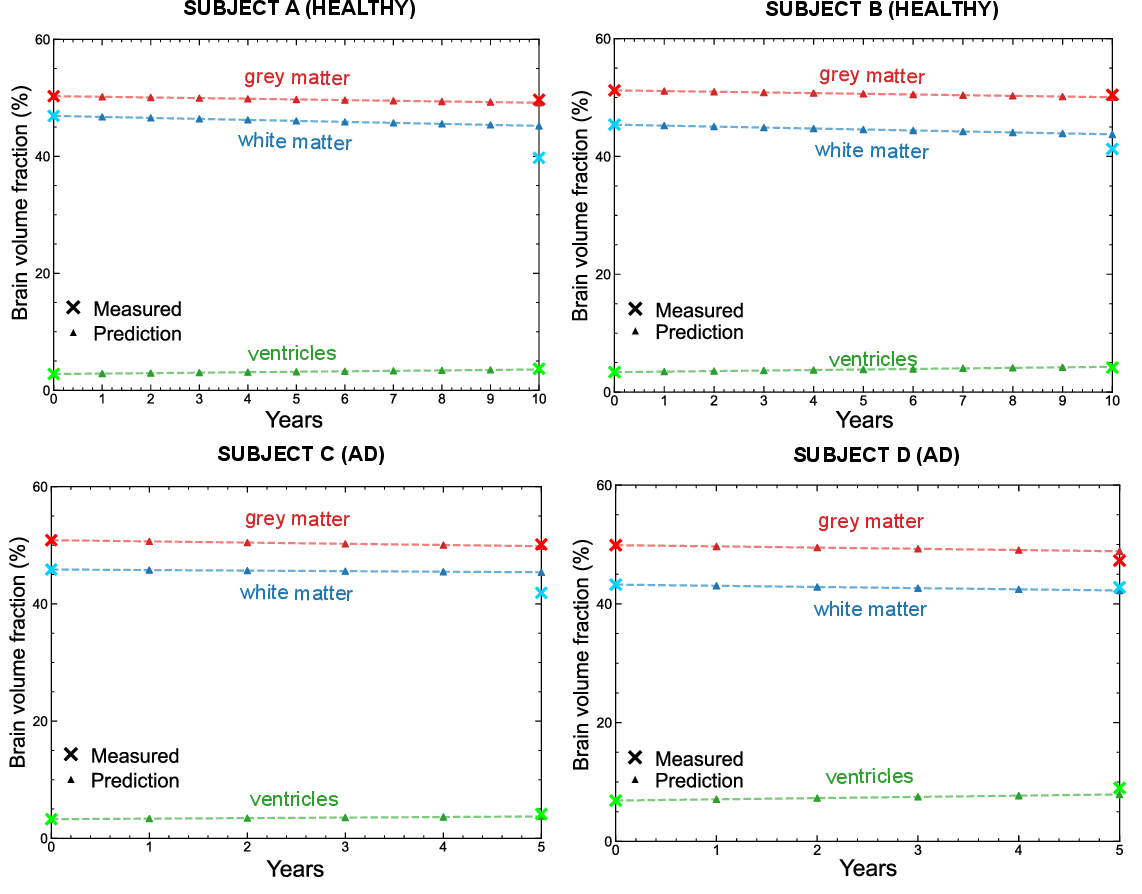}
    \caption{Temporal evolution of brain volume fractions for ventricles, white matter, and grey matter during the validation period. Predicted trajectories are compared with measured values at the final observation time for both healthy ageing and Alzheimer’s disease subjects.}
    \label{fig:validation_temporal}
\end{figure}

The brain volume fractions of the ventricles, white matter, and grey matter predicted for the considered subjects are shown in Fig. \ref{fig:validation_volumes}. For each subject, we compare the predicted values at the end of the simulation with the corresponding volumetric measurements obtained from medical imaging. The temporal evolutions of the brain volume fractions over the considered time horizons are presented in Fig. \ref{fig:validation_temporal}, for each subject. 
\\

Overall, the predicted volumetric fractions show good agreement with the measured data for both healthy ageing and Alzheimer’s disease cases. In the grey matter, the relative error between predictions and measurements was below 3.5\% for all subjects, healthy and undergoing Alzheimer's disease, with the relative error in the healthy ones being below or around 1\%. The predicted ventricular brain volume fractions are also in good agreement for the healthy subjects (relative error below 4\%), although higher discrepancies are attained for the AD subjects. Minor discrepancies are observed in the white matter fraction, which likely reflect modelling simplifications. White matter degeneration is known to be strongly influenced by microstructural processes such as axonal loss, which are not explicitly represented in the present framework and are instead captured through atrophy parameters. In addition, the assumption of almost homogeneous material properties and uniform atrophy rates may contribute to an underestimation of white matter atrophy. It is also worth noting that a large number of distinct white matter regions are grouped into a single white matter domain in the present analysis. This grouping averages the response over the entire white matter, which may mask degeneration occurring only in specific domains. A finer subdivision of white matter regions could help capture these localised effects and improve agreement with imaging-derived measurements \cite{tueni2026region}. Despite these limitations, the overall agreement supports the ability of the proposed model to reproduce subject-specific volumetric trends. The analysis could be further enhanced by including biomarker measurements that would allow direct comparisons between predicted toxic protein concentration and clinical data \cite{JackHoltzman2013}.





\section{Discussion}
\label{Sec:Discussion}

The results presented are now discussed in relation to existing models in the literature. The comparison focuses on three representative models that address, to varying degrees, the same phenomena considered. Table \ref{tab:discussion} summarizes the main differences between the proposed framework and previously published models.\\

Schäfer et al. \cite{Schafer2019CMA} couple protein propagation with tissue atrophy, reproducing characteristic spatial patterns of tau accumulation and brain volume loss. Their model, however, progresses more rapidly than ours, as the state reached in their simulations around year 8 corresponds approximately to year 12 in the present model, which appears more consistent with the slower progression reported in clinical studies \cite{JackHoltzman2013}. The spatial patterns are similar in both cases. A notable limitation of their framework is that it is implemented on two-dimensional brain slices, with fibre directions prescribed manually rather than reconstructed from diffusion tensor imaging, and without an explicit representation of the cerebrospinal fluid, which prevents the model from capturing ventricular enlargement, a well-known hallmark of neurodegeneration \cite{Apostolova2012}.\\

Corti et al. \cite{corti2023discontinuous} incorporate the full diffusion tensor into the formulation, and the spatiotemporal protein evolution obtained in their framework agrees well with the patterns observed in our model. Their model, however, is restricted to two dimensions, does not account for the interaction between tau and amyloid-$\beta$, whose influence on disease progression is widely accepted \cite{Roda2022Crosstalk}, and does not capture structural consequences of neurodegeneration such as ventricular enlargement.\\

Blinkouskaya et al. \cite{Blinkouskaya2021Frontiers} extend the problem to three dimensions, bringing their approach closest to the one adopted here. The axial patterns are broadly similar in both models, although the present framework predicts a higher concentration of pathology in the inferior temporal and medial temporal regions, consistent with clinical observations \cite{Braak1991}. Ventricular expansion occurs more rapidly in our simulations and agrees better with clinical data \cite{Apostolova2012}. An additional advantage of our framework is the explicit representation of a larger number of anatomical regions, which allows the response of individual brain structures to be analysed separately.\\

\begin{table}[h]
\centering
\caption{Comparison of modelling features across different studies.}
\begin{tabular}{lcccc}\toprule

\textbf{Model}& \textbf{3D} & \textbf{DTI anisotropy} & \textbf{Tau-A$\bm\beta$ interaction} & \textbf{Atrophy} \\
\midrule

Schäfer et al. \cite{Schafer2019CMA}       & $\times$ & Approx. & $\times$ & $\checkmark$ \\
Corti et al. \cite{corti2023discontinuous}         & $\times$ & $\checkmark$ & $\times$ & $\times$ \\
Blinkouskaya et al. \cite{Blinkouskaya2021Frontiers}  & $\checkmark$ & $\times$ & $\times$ & $\checkmark$ \\
\textbf{This work} & $\checkmark$ & $\checkmark$ & $\checkmark$ & $\checkmark$ \\

\hline
\end{tabular}
\label{tab:discussion}
\end{table}

\section{Conclusions}
\label{Sec:Concluding remarks}

In this work, we have presented a three-dimensional bio-chemo-mechanical computational framework to study the progression of Alzheimer’s disease. The proposed model integrates reaction-diffusion formulations for toxic tau and toxic amyloid-$\beta$ propagation with a continuum mechanics description of tissue degeneration. New modelling features are presented to handle the coupling between tau and amyloid-$\beta$ proteins, the interplay between a multi-protein system with bio-mechanical atrophy, and the rigorous treatment of anisotropic protein transport through axonal vector fields. The numerical results obtained show that these features have a significant influence on predictions. The work also brings methodological contributions, developing a fully automated preprocessing pipeline, termed \texttt{BrainImage2Mesh}, which generates subject-specific finite element brain meshes and corresponding axonal orientation fields directly from medical imaging data. This tool enables anatomically accurate geometric modelling and provides the structural information required for anisotropic transport simulations.\\

Numerical simulations performed on subject-specific brain geometries demonstrate that the proposed framework reproduces key hallmarks of Alzheimer’s disease, including localised tau accumulation, heterogeneous atrophy patterns, ventricular enlargement, and differential volume loss in grey and white matter. Cohort-level simulations further show consistent deformation trends across subjects and highlight the influence of individual brain geometry on disease progression.\\

In addition, the predictive capabilities of the proposed framework are assessed through clinical benchmarking against longitudinal imaging data. The model shows good agreement with measured regional brain volume changes, supporting its ability to capture morphological evolution.\\

The proposed framework offers a practical and extensible approach to modelling Alzheimer’s disease progression, and its two main constituents, the automated \texttt{BrainImage2Mesh} pipeline and the bio-chemo-mechanical finite element model, are released openly to the community and are available to download at \url{https://mechmat.web.ox.ac.uk/codes}.

\FloatBarrier
\section*{Acknowledgments}
\label{Acknowledge of funding}

\noindent The authors acknowledge the Instituto Tecnológico de Asturias (IUTA) for financial support through the grant SV-25-GIJON-01-25. E. Mart\'{\i}nez-Pa\~neda acknowledges financial support from UKRI's Future Leaders Fellowship programme [grant MR/V024124/1].
Data used in preparation of this article were obtained from the Alzheimer's Disease Neuroimaging Initiative (ADNI) database (adni.loni.usc.edu). As such, the investigators within the ADNI contributed to the design and implementation of ADNI and/or provided data but did not participate in analysis or writing of this report. A complete listing of ADNI investigators can be found at:  \url{http://adni.loni.usc.edu/wp-content/uploads/how_to_apply/ADNI_Acknowledgement_List.pdf}. ADNI is funded by the National Institutes of Health Grant U01 U19AG024904.




\end{document}